\documentclass[aps,english,twocolumn,groupedaddress,prb]{revtex4-1}

\usepackage{graphicx}
\usepackage{amssymb}
\usepackage{amsmath}
\usepackage[dvipsnames]{mathtools,xcolor}
\usepackage[colorlinks=true,
            linkcolor=BrickRed,
            citecolor=Violet,
            urlcolor=Violet]{hyperref}

\def\Tr{{\rm Tr}}
\newcommand{\bra}[1]{\langle\,#1\,|}
\newcommand{\ket}[1]{|\,#1\,\rangle}
\DeclareMathOperator*{\simc}{\sim}

\begin{document}

\title{Quantum circuit at criticality}

\author{Nicolas Mac\'e}
\affiliation{Laboratoire de Physique Th\'eorique, IRSAMC, Universit\'e de Toulouse, CNRS, UPS, France}

\begin{abstract}
    We study a simple quantum circuit model which, without need for fine tuning, is very close to sitting at the transition between ergodic and many-body localized (MBL) phases. 
    We probe the properties of the model on large finite-size systems, using a matrix-free exact diagonalization method that takes advantage of the shallow nature of the circuit.
    Moreover, we provide a qualitative entanglement bottleneck picture to account for the close-to-critical nature of the model.
\end{abstract}
\date{\today}
\maketitle
\section{Introduction}
Much effort has been recently devoted to studying the dynamics of isolated quantum systems.
In the strongly interacting regime, they are expected to thermalize\cite{rigol_thermalization}, and their eigenstates to verify the Eigenstate Thermalization Hypothesis (ETH)\cite{deutsch,srednicki}.
However, this picture fails in the Many-Body Localized (MBL) phase\cite{basko,gornyi_mirlin_polyakov,oganesyan_huse,znidaric_prosen,pal_huse} (see Refs.\ \onlinecite{nandkishore_huse,alet_laflorencie,abanin_altman_bloch_serbyn} for reviews): the ETH is violated and the system never thermalize.
Although the last decade has witnessed a surge of experimental \cite{schreiber,bordia_coupling,choi} and theoretical \cite{vosk_rg,potter_rg,thiery_rg,imbrie,luitz_laflorencie_alet,barlev_cohen_reichman,serbyn_papic_abanin,bardarson_jens_pollmann_moore} studies on the MBL problem, a complete understanding of the neighborhood of the ETH-MBL transition point is still lacking.
In particular, approaching the transition from the ETH side, an anomalous slowing down of the dynamics has been reported
\cite{agarwal_subdiffusion,znidaric_subdiffusion,
znidaric_subdiffusion,khait_subdiffusion,luitz_subdiffusion}.
Whether it genuinely signals the onset of a ``bad metal'' phase or is a finite-size effect \cite{serbyn_thouless,lezama_slow_dynamics} remains to be clarified (see Refs.\ \onlinecite{luitz_barlev_review,agarwal_subdiffusion_review} for reviews).
Morevover, the gap remains to be bridged between the phenomenological Renormalization Group (RG) predictions for the transition\cite{vosk_rg,potter_rg,thiery_rg,goremykina_rg,morningstar_rg} and the numerical and experimental observations\cite{schiro_tarzia}.
To address these issues, and more generally to better understand the features of the transition, one could in principle numerically approach it to a higher precision.
However, that would requires probing larger times and length scales, a challenge for present day numerical methods.

Quantum circuits offer a promising framework for capturing the essential features of the ETH-MBL transition.
In these simple toy models, the state of the system evolves in time under the action of local unitary gates.
Periodically driven Floquet quantum circuits can host both thermal and MBL phases\cite{zhang_floquet_mbl, abanin_theory_floquet, yao_time_crystal, lezama_slow_dynamics}.
Their features are qualitatively similar to the ones observed in generic Floquet MBL systems \cite{ponte_noncircuit_floquet, lazarides_noncircuit_floquet, bordia_floquet_experiment, else_time_crystal, khemani_time_crystal}.
Furthermore, quantum circuits are easier to handle than convential many-body Hamiltonian models.
This has notably lead to the understanding of their entanglement dynamics \cite{znidaric_entanglement, oliveira_entanglement, gutschow_entanglement_clifford, nahum_entanglement_growth, nahum_entanglement_quenched, vonkeyserlingk_entanglement_OTOC, nahum_operator_entanglement, rakovszky_subballistic_entanglement, huang_subballistic_entanglement, rakovszky_inhomogenous_quench} and information scrambling properties \cite{nahum_operator_spreading, vonkeyserlingk_entanglement_OTOC, roberts_frame_potential_otoc, rakovszky_otoc_conserved} in the fully ergodic regime.
Encouraged by this success, we propose here that a simple Floquet quantum circuit can be used as a model of the ETH-MBL transition which is both (i) realistic (including in particular quantum effects that are absent from the phenomenological RG pictures) and (ii) simple enough to be amenable to exact treatment.

We demonstrate the relevance of the model for the study of the MBL problem by extensive numerical simulations, employing an iterative exact diagonalization method that takes advantage of the shallow quantum circuit structure of the Floquet operator.
Moreover, we numerically show that the system is close to being at the critical point in the \emph{uniform} limit (which we define in detail later).
Once more taking advantage of the simple quantum circuit nature of the model, we study the close-to-critical properties of the uniform limit, and propose in particular a phenomenological entanglement bottleneck picture.

The remainder of the article is organized in five parts.
After introducing the model in Sec.\ \ref{sec:model}, we numerically discuss its properties, and in particular its critical nature, in Sec.\ \ref{sec:criticality}.
Sec.\ \ref{sec:bottlenecks} presents the entanglement bottleneck picture, and discusses how it is tied to the anomalous dynamics and the close-to-critical nature of the model.
In Sec.\ \ref{sec:perturbation}, we study how altering the model drives it away from criticality.
Finally, Sec.\ \ref{sec:conclusion} gathers our conclusions.
\section{Model}
\label{sec:model}
\begin{figure}[htp]
\centering
\includegraphics[width=\columnwidth]{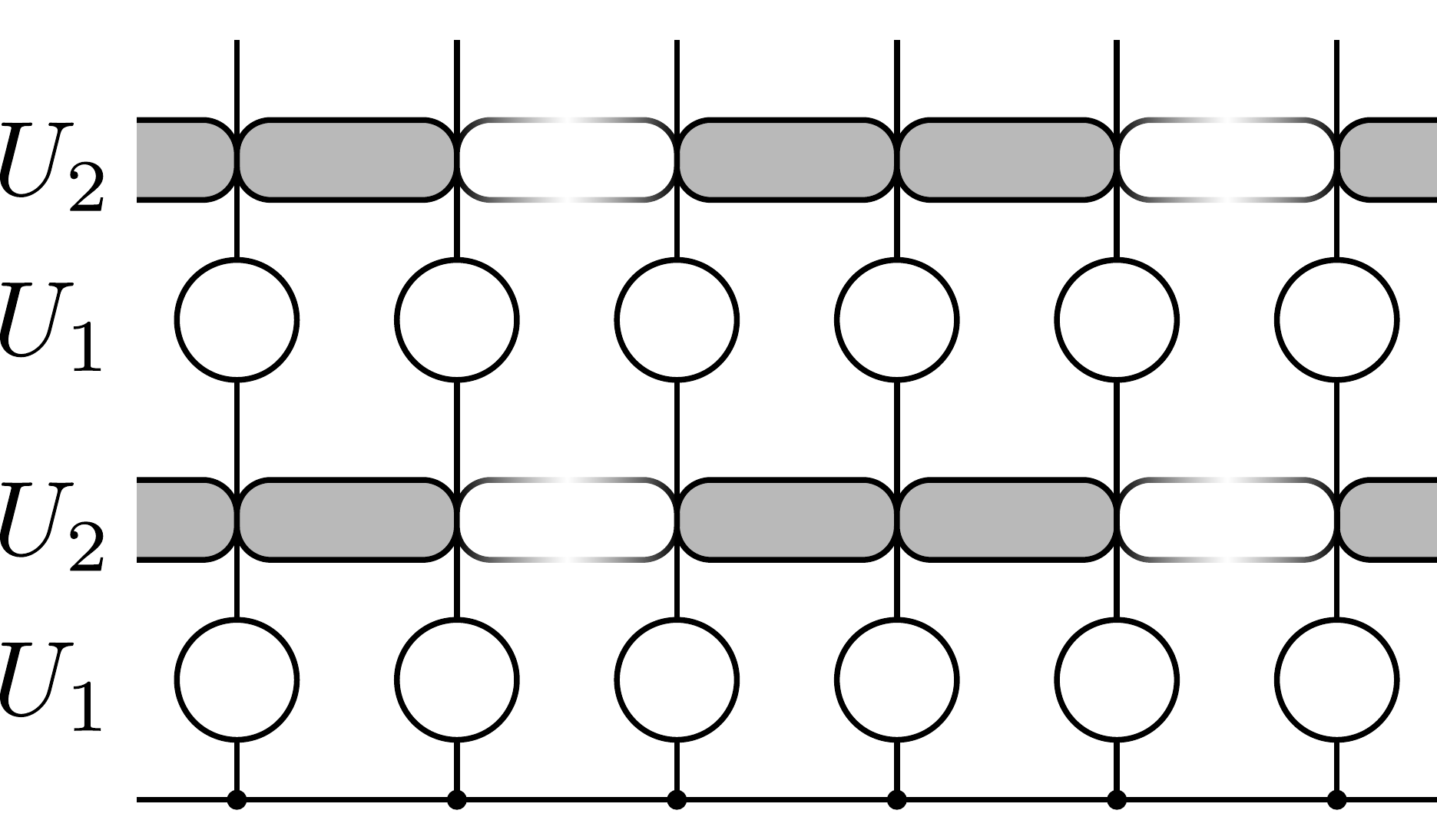}
\caption{Space-time representation of the Floquet circuit. The ``scrambling'' operator $U_1$ consists of 1-gates drawn from the Haar distribution, while the ``dephasing'' operator $U_2$ consists of random diagonal 2-gates. Dephasing gates are colored in white at places where a bottleneck almost cuts the system in two halves. Bottlenecks drive the critical physics of the model, as we argue.}
\label{fig:model}
\end{figure}

We study here a Floquet quantum circuit first introduced in Ref.\ \onlinecite{chan_dephasing}.
As shown on Fig.\ \ref{fig:model}, the Floquet operator is composed of two layers: $U = U_2 \times U_1$.
The first layer is a scrambling layer of 1-qubit random Haar unitaries:
\begin{equation}
	U_1 = \prod_{j=1}^{L} e^{-i \sum_{\alpha=0}^{3} h_j^\alpha \sigma_j^\alpha},
\end{equation}
with $\sigma_j^0$ the identity, $\sigma_j^{\alpha > 0}$ the Pauli operators acting on qubit at site $j$ and $L$ the total number of qubits.
The $h_j^\alpha$ are chosen so that the local operators are distributed according to the Haar measure.
The second layer is a dephasing layer:
\begin{equation}
\label{eq:dephasing_layer}
	U_2 = \prod_{j=1}^{L-1} e^{-i J_j \sigma_j^z  \sigma_{j+1}^z}  \prod_{j=1}^{L}  e^{-i  h_j \sigma_j^z}.
\end{equation}
$J_j$ and $h_j$ are Gaussian random variables of zero mean and standard deviation $J$:
\begin{align*}
	\overline{J_j} &= \overline{h_j} = 0,\\
	\overline{J_j^2}&=\overline{h_j^2} = J^2.
\end{align*}

This model belongs to the extensively studied family of binary drives alternating between dephasing ($\sigma^z$ gates) and scrambling ($\sigma^x$ and $\sigma^y$ gates) layers, sometimes termed ``Kicked Ising'' models.
They are usually observed to host ETH and MBL phases \cite{zhang_floquet_mbl, abanin_theory_floquet, yao_time_crystal, lezama_slow_dynamics}.
In the model at hand, the only free parameter is the dephasing strength $J$.
If $J=0$, we have $U_2=1$ and no dephasing occurs.
Entanglement production is therefore arrested, and the system is trivially localized.
As $J$ is increased, more and more relative time is spent in the dephasing layer, increasing the entanglement production of the system.
Whether or not it becomes ETH when $J$ is increased is not obvious, and Ref.\ \onlinecite{chan_dephasing} argued from numerical observations that the model should stay MBL, even in the $J \to \infty$ limit.
However, we observe a slow flow towards ergodicity with increasing system size, indicating that the system will eventually become ETH for $J$ large enough.
The slowness of the flow signals a model close to being at the critical point between ETH and MBL phases.
Since when $J \to \infty$ the dephasings (Eq.\ \ref{eq:dephasing_layer}) are uniformly distributed, we call this close-to-criticality model \emph{uniform}.
\section{Criticality}
\label{sec:criticality}
\begin{figure}[htp]
\centering
\includegraphics[width=1\columnwidth]{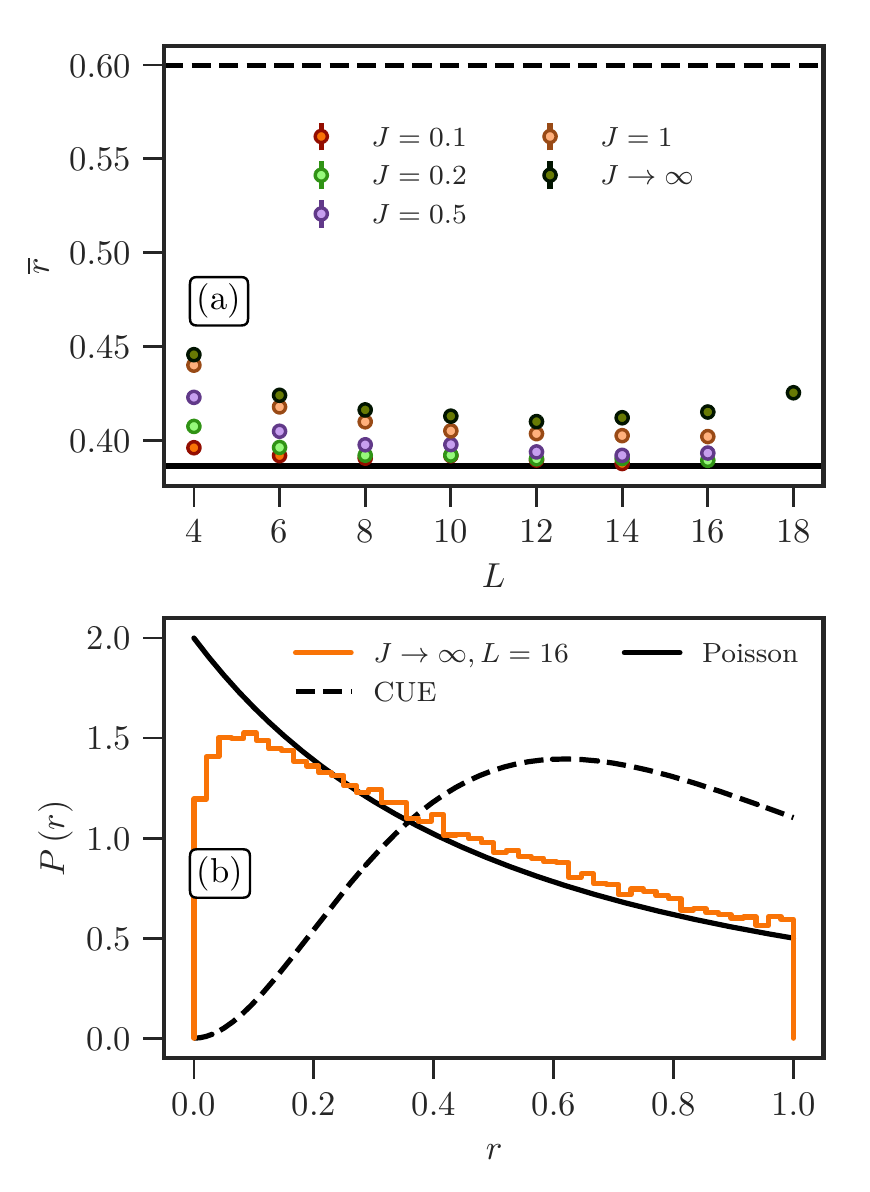}
\caption{(a) Average gap ratio as a function of system size, for different dephasing strength $J$. Average is performed over at least 3000 samples. Errorbars are smaller than the symbols. Solid line: Poisson gap ratio, dashed line: CUE gap ratio. (b) Gap ratio distribution in the uniform ($J \to \infty$) case. Solid line: Poisson gap ratio distribution, dashed line: CUE distribution. We approximate the CUE distribution by the GUE surmise of Ref.\ \onlinecite{atas_rgaps}, which is very close to the CUE result \cite{dalessio_rgaps_floquet}.}
\label{fig:rgaps_dephasing}
\end{figure}
In this part we focus on the uniform model with $J \to \infty$, which is close to being critical, as we discuss now.

\emph{Level statistics ---}
The statistics of energy levels is an effective probe of the thermal or localized nature of Hamiltonian systems\cite{oganesyan_huse}.
The same analysis can be carried out in the case of a Floquet system, trading the energies for the quasienergies $\theta$, defined as
$$
	U = \sum_\theta e^{i\theta} \ket{\theta}\bra{\theta},
$$
where $\{\ket{\theta}\}$ is the complete basis of Floquet eigenstates.
Ordering the quasienergies $\theta_{n} < \theta_{n+1} < \dots$, the gap ratio $r_n$ is defined as the normalized ratio of two consecutive spectral gaps: $r_n = \min\{g_{n+1}/g_n, g_{n}/g_{n+1}\}$, with $g_n = \theta_{n} - \theta_{n-1}$.
In an MBL phase, there is no level repulsion, and hence a Poisson-distributed spectral statistics is observed.
One has $r_\text{Poisson} = 2 \ln 2 - 1 \simeq 0.38$.
In a thermal phase, the system's spectrum coincides with the one of a random matrix, here drawn from the circular unitary ensemble (CUE).
One has $r_\text{CUE} \simeq 0.60$\cite{dalessio_rgaps_floquet}.

We have numerically computed the gap ratios, extracting at least three consecutive energy levels from the spectrum.
Since energy is not conserved, it does not matter where in the quasienergy spectrum we compute the level statistics.
At zero dephasing $J = 0$, the system is trivially integrable, and we have $\overline r = r_\text{Poisson}$.
For small enough dephasing, we therefore expect  the system to be in an MBL phase.
Supporting this hypothesis, we indeed observe $\overline r = r_\text{Poisson}$ for $J \lesssim 1$ (see Fig.\ \eqref{fig:rgaps_dephasing} (a)).
At larger dephasing strength, the average gap ratio crosses over to a larger value, a signature of the integrability breakdown associated to delocalization.
The most delocalized model is the uniform one, where $J \to \infty$.
Interestingly, in this case we observe a non-monotonous flow: the gap ratio initially flows down towards Poisson, then meets a turning point at $L=12$, and subsequentially flows up, towards the CUE value.
Such a non-monotonous behavior is numerically observed on the ergodic side of the MBL transition, very close to the critical point.
It is e.g.\ responsible for the slow drift in the crossing of various physical observables\cite{oganesyan_huse, kjall_entropy, luitz_laflorencie_alet}.
The same phenomenon is observed in phenomenological RG models of the transition\cite{potter_rg, morningstar_rg}, where the predicted Kosterlitz-Thouless-like scaling dictates that on the ergodic side close to the transition, the system initially flows towards localization, before eventually thermalizing.
Therefore, the flow observed in the uniform model is fully consistent with the eventual thermalization of the system.

Let us stress here that the thermalizing trend is visible thanks to a specially tailored exact diagonalization method allowing us to reach $L=18$.
On smaller system sizes $L \leq 14$ accessible to standard ED methods, it is difficult to argue for the thermal nature of the uniform model.
Even for the largest system of $L=18$ qubits, the average gap ratio is very far from the CUE value we expect it will reach eventually, as seen on Fig.\ \ref{fig:rgaps_dephasing} (a).
Examining the gap ratio distribution (Fig.~\ref{fig:rgaps_dephasing} (b)) further confirms that the system is both far from the localized (Poisson) and ergodic (CUE) limits.
This indicates that the thermalization length scale $L_\mathrm{th}$, defined as the length scale above which the system is reasonably thermalized, is extremely large.
This in turn indicates that the uniform model is very close to the critical point.

\begin{figure}[htp]
\centering
\includegraphics[width=1\columnwidth]{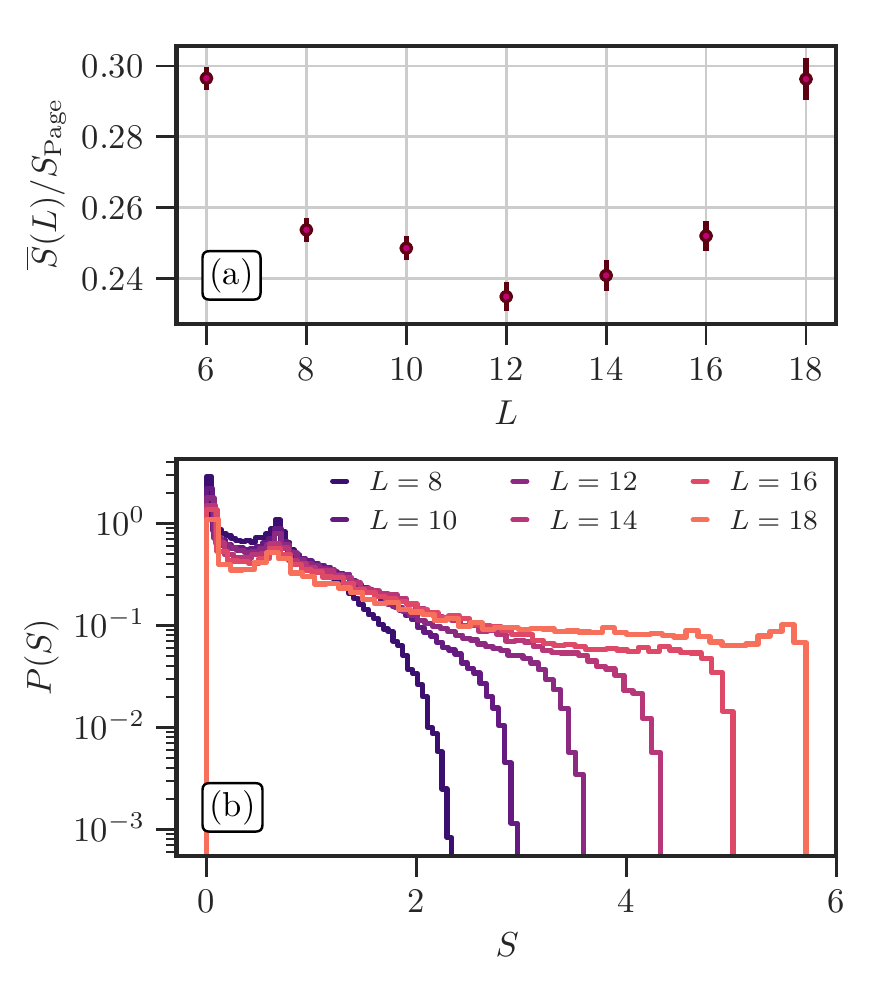}
\caption{(a) Average half-chain von Neumann entanglement entropy divided by the Page value, as a function of system size. At least 3000 samples are used. (b) Entanglement entropy distribution as a function of system size for the uniform ($J \to \infty$) model.}
\label{fig:entropy_critical}
\end{figure}
\emph{Eigenstate entanglement ---}
To measure the eigenstate localization degree, we compute the von Neumann entanglement entropy.
Given a bipartition of the system into regions $A$ and $B$, it is defined as $S(\psi) = - \text{Tr}(\rho_A \ln \rho_A)$, with $\rho_A = \Tr_B \ket{\psi}\bra{\psi}$ the reduced density matrix of the state $\psi$.
If $\psi$ is localized in region $A$, there is almost no entanglement between $A$ and $B$, and the entropy $S(\psi)$ is close to 0.
However, if $\psi$ is extended over $A$ and $B$, we expect the entanglement entropy to be large and to scale as the volume of the region: $S(\psi) \sim |A|$.
We consider here the half-chain bipartition $L_A = L/2$.
In an ETH phase and in the large size limit, the entanglement entropy is a Gaussian centred on the Page value\cite{page} $S_\text{Page} = \ln 2 \times L/2 - 1/2$.
In a localized phase, the half-chain entanglement entropy is sub-volumic\cite{bauer_nayak}, \emph{i.e.} the entanglement entropy density vanishes: $S_\text{loc}/S_\text{Page} \to 0$.

We have examined (Fig.\ \ref{fig:entropy_critical} (a)) how the entanglement entropy divided by the Page value depends on system size, in the uniform model.
We observe the exact same behavior as for the gap ratio: it first flows down, then inflects at $L \simeq 12$ to flow up, indicating that increasing system size ultimately drives it towards the thermal phase.
We expect the entropy to eventually reach the Page value, from which we are quite far when $L \leq 18$.
This again indicates that the thermalization length scale $L_\mathrm{th}$ is large.

Fig.\ \ref{fig:entropy_critical} (b) shows the distribution of entanglement entropies in the uniform model.
For small system sizes $L \leq 12$, the distribution has a sharp maximum at 0, and decays rapidly. A secondary peak can be observed at $\ln 2$. 
All these features are characteristic of the MBL phase \cite{lim_sheng,luitz_entropy_tails}.
As system size is increased, the 0 and $\ln 2$ peaks are suppressed, weight is transferred to larger entropy, and the distribution becomes broader and broader.
The broadening of the entropy distribution signals the proximity of a phase transition.
It has in particular been observed at the ETH-MBL critical point, where the entropy variance is maximal \cite{luitz_laflorencie_alet,kjall_entropy}.
This observation therefore further confirms that the uniform model is very close to being critical.

Such a close to critical system is interesting, in particular because the system size can be simultaneously large compared to the microscopic length scale and small compared to the thermal length scale: $1 \ll L \ll L_\mathrm{th}$.
In that critical regime, the entropy distribution approximates well the one at the MBL critical point.
Observing on Fig.\ \ref{fig:dephasing_gate_entanglement} (b) that the distribution becomes flatter and flatter with increasing system size, we postulate that the half-chain entanglement entropy could be uniformly distributed at the transition.
This claim is further supported by previous works on Hamiltonian models\cite{lim_sheng,luitz_entropy_tails}, where a flattening of the entropy distribution is observed close to the transition.
\section{Bottleneck picture}
\label{sec:bottlenecks}
\begin{figure}[h]
\centering{
   \fontsize{14pt}{13pt}\selectfont
   \def\svgwidth{3.333in}
\resizebox{60mm}{!}{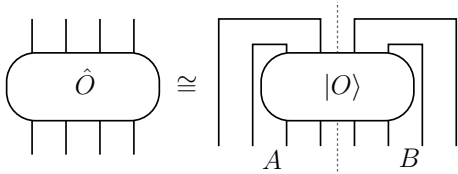}
\caption{Illustration of the reshape operation mapping an operator $\hat O$ (left) to a vector $\ket{O}$ on the doubled Hilbert space (right). One can then define an $A,B$ bipartition with respect to which the operator entanglement entropy is computed.}
\label{fig:operator_reshape}
}
\end{figure}
In the neighborhood of the transition, sub-ballistic information dynamics and sub-diffusive charge transport is observed on finite-size systems\cite{agarwal_subdiffusion,agarwal_subdiffusion_review,khait_subdiffusion,znidaric_subdiffusion,luitz_subdiffusion,sahu_scrambling}.
This has been attributed to rare MBL-like ``Griffiths'' regions, which act as bottlenecks for the dynamics.
In this section, we exploit the simple quantum circuit nature of our model to determine the microscopic bottleneck distribution.
We discuss how it relates to the numerically observed sub-ballistic entanglement growth, and discuss the possible connection with the close-to-critical nature of the uniform model.

\emph{Operator entanglement entropy ---}
We call ``microscopic bottlenecks" the links $(j,j+1)$ where entanglement production is strongly suppressed.
To quantify bottleneck strength, we employ the operator entanglement entropy of the evolution operator $U$.
As illustrated on Fig.\ \ref{fig:operator_reshape}, an operator $\hat O$ acting on $\mathcal{H}$ can be viewed as a vector $\ket{O}$ acting on the doubled Hilbert space $\mathcal{H} \otimes \mathcal{H}$. The operator entanglement entropy of $\hat O$ is defined as the entanglement entropy of $\ket{O}$.
Roughly speaking, the operator entanglement entropy for the $A,B$ bipartition (see Fig.\ \ref{fig:operator_reshape}) quantifies how far an operator $\hat O$ is from being written as the tensor product $\hat O_A \otimes \hat O_B$.
In other words, the larger the operator entanglement entropy is, the more the action of $\hat O$ on a state generates entanglement between $A$ and $B$.
 
\emph{Microscopic bottlenecks ---}
 Along the lines of Ref.\ \onlinecite{nahum_entanglement_quenched}, we identify ``weak links'' acting as bottlenecks.
 An obvious way to create an infinitely weak link at position $(j,j+1)$ is to set $J_j = 0$.
Then, the evolution operator writes as a tensor product $U = U_\text{left} \otimes U_\text{right}$ with $U_\text{left}$ (resp.\ $U_\text{right}$) acting only on the degrees of freedom of the left (resp.\ on the right) of the qubit at position $j$.
In other words, the chain is disconnected, and no entanglement is generated through the $(j,j+1)$ link.
This is a ``infinitely strong'' bottleneck.
To quantify how strong a finite bottleneck with $J_j \ll 1$ is, we compute the operator entanglement entropy of the evolution operator $U$ at the $(j,j+1)$ link.
It is easily computed, thanks to the shallow quantum circuit nature of the operator.
It is equal to the entropy of the dephasing gate $\Phi = \exp(i J_j \sigma^z_j \sigma^z_{j+1})$ acting on qubits $j$ and $j+1$.
We find
\begin{equation}
\label{eq:dephasing_gate_entanglement}
	S_{\Phi}(j|j+1) = -\cos^2 J_j \ln\left(\cos^2 J_j\right) -\sin^2 J_j \ln\left(\sin^2 J_j\right). 
\end{equation}
Interestingly, this coincides with the maximal amount of entanglement entropy one can generate by applying the evolution operator on a disentangled state \cite{kraus_cirac}.
When $J \ll 1$, operator entanglement is close to zero: $S \sim -J^2\ln J^2 \ll 1$.
Since the number of channels that can be entangled after $t$ applications of the Floquet operator is proportional to $t$, the entanglement across a bottleneck grows linearly with time, with a bottleneck-dependent velocity $v_E = -J^2\ln J^2$:
\begin{equation}
\label{eq:op_ent_growth}
	S_{U(t)}(j|j+1) \simc_{t \ll 1/J^2} -J^2\ln J^2\times t.
\end{equation}
The same form was obtained in Ref.\ \onlinecite{nahum_entanglement_quenched} for the entropy growth of initial states.
This comes as no surprise, since operator and state entropy are expected to behave in a qualitatively similar way\cite{lezema_luitz}.
Eq.\ \eqref{eq:op_ent_growth} shows that $S_{\Phi}$ is equal to the rate at which entanglement entropy is generated at the bottleneck. Therefore, it makes up for a natural measure of the bottleneck strength.

\emph{Bottlenecks in the uniform model ---}
\begin{figure}[htp]
\centering
\includegraphics[width=1\columnwidth]{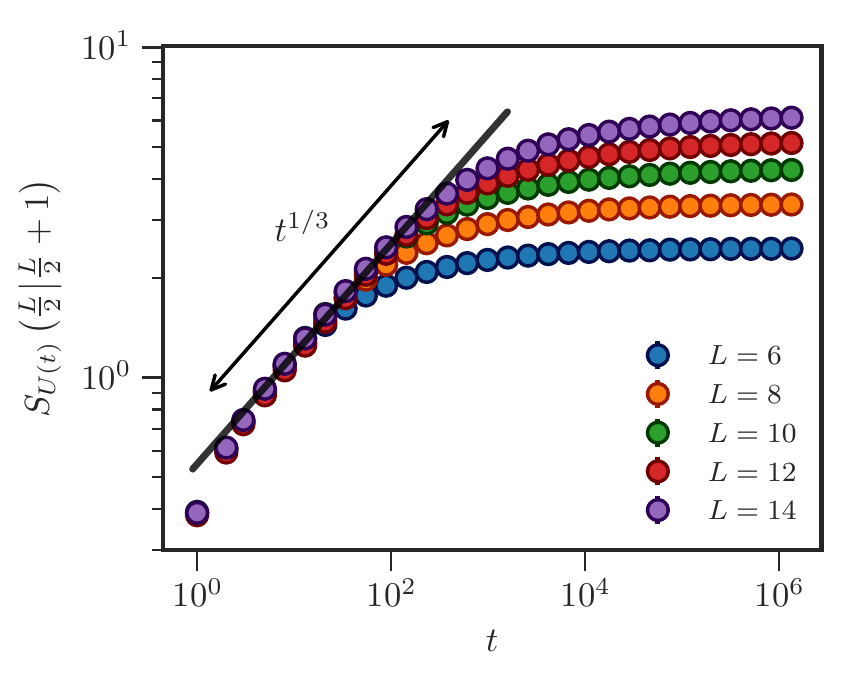}
\caption{Half-chain operator entanglement entropy as a function of time. Error bars are smaller than the size of the symbols. After a short-time regime, the entropy growth is well fitted by a power-law (straight line) with a $1/3$ exponent related to the power-law tail of the bottleneck distribution (see main text).}
\label{fig:dephasing_gate_entanglement}
\end{figure}
In the uniform model, we obtain for the operator entanglement distribution:
\begin{equation}
\label{eq:rate_distribution}
	P\left(S_{\Phi}\right) \simc_{S_\Phi \ll 1} \frac{2}{\pi \sqrt{|\ln S_{\Phi}|}} \frac{1}{\sqrt{S_{\Phi}}}.
\end{equation}
Crucially, the distribution diverges as $S_\Phi \to 0$.
This has been predicted \cite{nahum_entanglement_quenched} to lead to anomalous transport and operator dynamics.
Moreover, the entanglement entropy is expected to grow sub-ballistically:
\begin{equation}
	\overline{S}(t) \sim S_0 t^{1/z_s},
\end{equation}
with a dynamical exponent $z_s = (a+2)/(a+1)$, where $a$ is the exponent of the local entanglement rate distribution, given in our case (Eq.\ \eqref{eq:rate_distribution}) by $a = -1/2$.
The predicted growth rate $S(t) \sim t^{1/3}$ is compatible with numerical data, as shown on Fig.\ \ref{fig:dephasing_gate_entanglement} (b).
This confirms the crucial role microscopic bottlenecks play in the dynamics of the system.

\emph{Renormalization of microscopic bottlenecks ---}
Considering now a chain of finite length $L$, extreme value statistics\cite{majumdar_extreme_value} predicts that on average, the weakest link decays as a power-law of system size:
$$
	S_\Phi^\text{min}(L) \simc_{L \to \infty} \frac{\ln(c L)}{(cL)^2},
$$
with $c=2/\pi$.
Therefore, the Floquet operator acting on $L$ qubits decomposes into two blocks that almost commute:
$$
	U(L) = U(L_b) \otimes U(L - L_b) + \mathcal{O}\left(\frac{1}{L^2}\right),
$$
where $L_b$ is the location of the strongest bottleneck.
While this means that the Floquet operator breaks into at least two disconnected pieces in the $L \to \infty$ limit, it does not entail arrested transport and localization.
Indeed, transport is prevented only if the infinite time evolution operator $\lim_{L \to \infty} \lim_{t \to \infty} U(t)$ breaks into disconnected parts.
However this is not the case for the uniform model: numerical computation of $S_{U(t)}$ shows that as time is increased, the strongest bottleneck of a chain of fixed length $L$ becomes weaker and weaker, and that the initially almost decoupled chain pieces are eventually connected, enabling transport and delocalization.
In other words, microscopic bottlenecks are renormalized at larger distances, driving the system towards ergodicity.
We think that a precise characterization of the bottleneck renormalization flow should be feasible in the uniform toy model, owing to its simplicity.
Moreover, it would lead to a better understanding of the close to critical nature of the mode.

\section{Breaking criticality}
\label{sec:perturbation}
In this section, we modify the model in two different ways, in order to remove bottlenecks.
In both cases, this destroys -- as expected -- the almost critical nature of the uniform model, and restores the full ETH phase.

\emph{Uniform qutrit model ---}
\begin{figure}[htp]
\centering
\includegraphics[width=1\columnwidth]{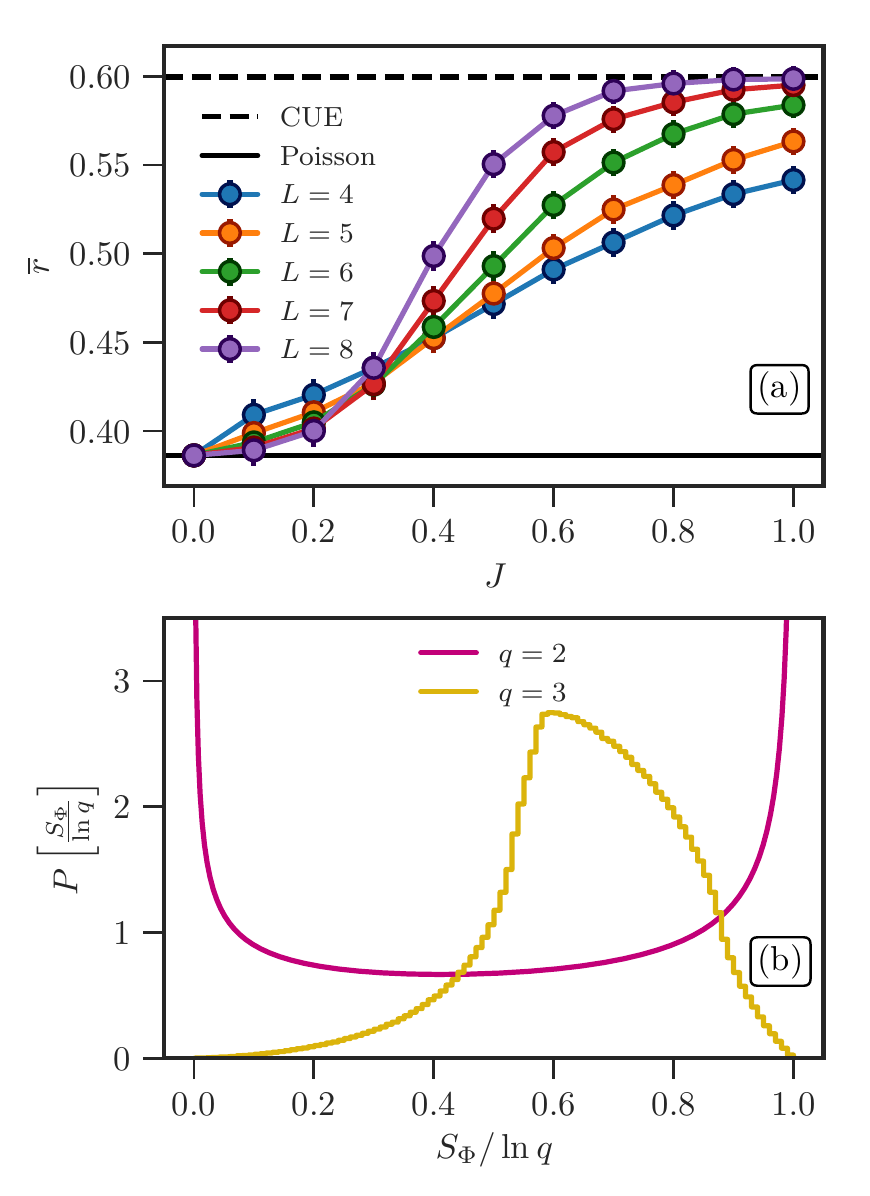}
\caption{(a) Average gap ratio in the qutrit model as a function of dephasing strength $J$, for different system sizes $L$. Average is performed over at least 3000 samples, using all of the $3^L$ eigenstates. (b) Phase gate entanglement distribution in the uniform $J \to \infty$ model, for qubits ($q=2$) and qutrits ($q=3$).}
\label{fig:d3}
\end{figure}
\begin{figure}[htp]
\centering
\includegraphics[width=1\columnwidth]{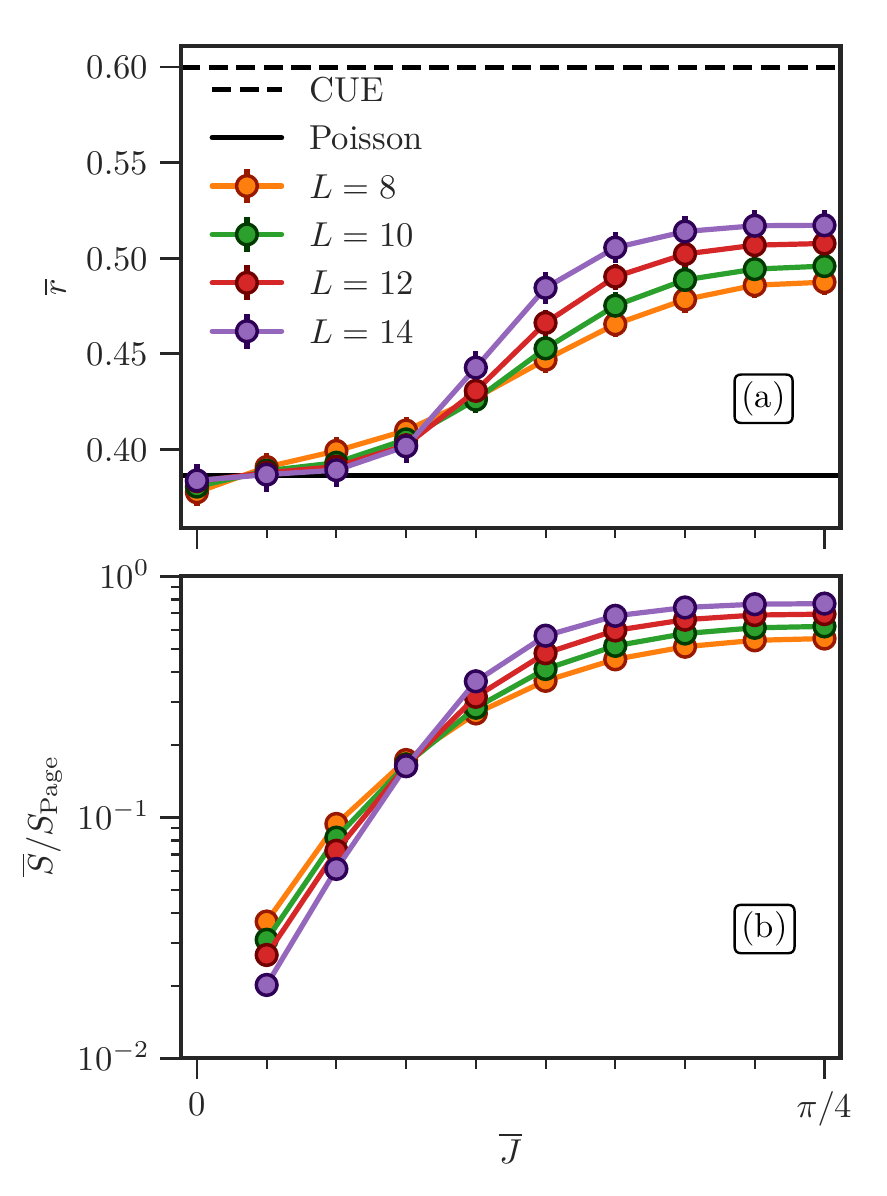}
\caption{Observables in the Kicked Ising model, as a function of average  dephasing $\overline{J}$. (a) Average gap ratios, (b) Average entanglement entropy divided by Page value. Data is averaged over all eigenstates and at least 1000 samples.}
\label{fig:KI_transition}
\end{figure}
We extend the model to the case where the local Hilbert space is of dimension $q = 3$ (qutrits) instead of $q=2$ (qubits).
Following Ref.\ \onlinecite{chan_dephasing}, we now consider a scrambling layer of random Haar 1-qutrit unitaries, followed by a dephasing layer of 2-qutrit gates.
As before, the dephasing gate entries are independent Gaussian random variables of zero mean and standard deviation $J$.
Finally, remark that the Hilbert space dimension is $3^L$ for a chain of $L$ qutrits, limiting full exact diagonalization to fairly small system sizes.
However, this does not prevent us from observing the onset of thermalization, as we discuss now.

When the dephasing strength $J$ is small, the gap ratio is close to the Poisson value (Fig.\ \ref{fig:d3} (a)). This indicates that the system is localized.
At larger $J$, instead of a slow flow towards ergodicity, we observe a fast one, with a clear transition point, as was already pointed out in Ref.\ \onlinecite{chan_dephasing}.
For $J \geq 0.9$, the spectral statistics of a system of only $L=8$ qutrits is consistent with full ergodicity. This is in contrast with the $q=2$ case, where even in the limit $J \to \infty$ and for systems as large as $L = 18$, ergodicity is far from being reached.

The difference between the $q=2$ and $q=3$ cases can be understood from the analysis of the dephasing gate entanglement entropy distribution (Fig.\ \ref{fig:d3} (b)).
In the qubit case, even when $J \to \infty$, the distribution diverges as a power-law at small entropy.
This means that microscopic bottlenecks drive the dynamics, leading, as discussed in the previous section, to sub-ballistic operator spreading, slow entanglement growth and slow thermalization.
By contrast, the distribution does not diverge in the qutrit case: bottlenecks are then weak enough for the operator spreading to be ballistic\cite{nahum_entanglement_growth} and for the entanglement growth and thermalization to be fast.
To gain an intuition of why this is the case, let us introduce the Kraus-Cirac (KC) number\cite{kraus_cirac,soeda_kc_number}, defined as the number of parameters involved in the operator entanglement of the dephasing gate.
Indeed, while the dephasing gate depends on $q$ parameters, its entropy, being invariant under local unitary transformations, is much more constrained.
On the one hand, Eq.~\eqref{eq:dephasing_gate_entanglement} shows that $\mathrm{KC}(q=2)=1$.
On the other hand, one can show\cite{chan_dephasing} that $\mathrm{KC}(q=3)=4$: there are many more parameters, and therefore many more ways for the qutrit dephasing gate to generate entanglement.
In the $J \to \infty$ limit, the parameters of the qutrit dephasing gate take all accessible values with equal probability, and microscopic bottlenecks are very rare in the qutrit case.

As we have argued in the previous section, bottlenecks are a crucial feature for the close-to-critical nature of the uniform qubits model.
In the absence of microscopic bottlenecks, there is no obstruction to the thermalization of the system on short length scales in the large $J$ limit, thus explaining the fast thermalization observed in the qutrit model.

\emph{Dephasing shift ---}
Another way to destroy criticality is to suppress bottlenecks directly in the $q=2$ model.
This can be achieved by increasing the average value of the dephasings, which was previously set to $\overline{J_j} = 0$.
Choosing a zero variance ($J=0$), and $\overline{J} > 0$, we see from Eq.\ \eqref{eq:dephasing_gate_entanglement} that microscopic bottlenecks are completely suppressed.
If $\overline{J}$ is large enough, we now expect the ETH predictions to be verified on systems of modest size.
Fig.\ \ref{fig:KI_transition} confirms our analysis: for $\overline{J} \lesssim 0.2$, the gap ratio converges to Poisson and the entanglement entropy is sub-extensive, as they should in an MBL phase, while for $\overline{J} \gtrsim 0.2$ the gap ratio and the entanglement entropy density quickly flow towards to their ETH value as system size is increased.
\section{Conclusion}
\label{sec:conclusion}
Our work demonstrates that a simple quantum circuit model can, without fine-tuning, stand close to the critical point between ETH and MBL phases.
The computation of the spectral statistics and eigenstate entanglement up to large system sizes, using an iterative method that exploits the shallow circuit nature of the model, shines light on the properties of the phase transition.
Furthermore, numerical observation of anomalous, sub-ballistic operator entanglement growth motivates a simple characterization of the Griffiths physics of the critical region in terms of microscopic bottlenecks, that yields a dynamical exponent in good agreement with numerical data.
The study of such simple quantum circuit models opens up the exciting perspective of bridging the gap between phenomenological classical models of the transition amenable to RG treatment, and the more realistic quantum microscopic models.

\section*{Acknowledgements}
I would like to thank Fabien Alet, Julia Baarck, Will Berdanier, Sam Garratt, Nicolas Laflorencie and Hugo Théveniaut for fruitful discussions.
I am especially grateful to Andrea de Luca for introducing me to the model and suggesting using an iterative diagonalization method.
This work benefited from the support of the French National Research Agency (ANR) (grant THERMOLOC ANR-16-CE30-0023-02) and the Programme Investissements d'Avenir (grant ANR-11-IDEX-0002-02, ANR-10-LABX-0037-NEXT).
I acknowledge the use of computing resources from CALMIP (grants No.\ 2017-P0677 and No.\ 2018-P0677)  and GENCI (grant No.\ x2018050225).
\bibliography{dephasing}

\begin{thebibliography}{67}%
\makeatletter
\providecommand \@ifxundefined [1]{%
 \@ifx{#1\undefined}
}%
\providecommand \@ifnum [1]{%
 \ifnum #1\expandafter \@firstoftwo
 \else \expandafter \@secondoftwo
 \fi
}%
\providecommand \@ifx [1]{%
 \ifx #1\expandafter \@firstoftwo
 \else \expandafter \@secondoftwo
 \fi
}%
\providecommand \natexlab [1]{#1}%
\providecommand \enquote  [1]{``#1''}%
\providecommand \bibnamefont  [1]{#1}%
\providecommand \bibfnamefont [1]{#1}%
\providecommand \citenamefont [1]{#1}%
\providecommand \href@noop [0]{\@secondoftwo}%
\providecommand \href [0]{\begingroup \@sanitize@url \@href}%
\providecommand \@href[1]{\@@startlink{#1}\@@href}%
\providecommand \@@href[1]{\endgroup#1\@@endlink}%
\providecommand \@sanitize@url [0]{\catcode `\\12\catcode `\$12\catcode
  `\&12\catcode `\#12\catcode `\^12\catcode `\_12\catcode `\%12\relax}%
\providecommand \@@startlink[1]{}%
\providecommand \@@endlink[0]{}%
\providecommand \url  [0]{\begingroup\@sanitize@url \@url }%
\providecommand \@url [1]{\endgroup\@href {#1}{\urlprefix }}%
\providecommand \urlprefix  [0]{URL }%
\providecommand \Eprint [0]{\href }%
\providecommand \doibase [0]{http://dx.doi.org/}%
\providecommand \selectlanguage [0]{\@gobble}%
\providecommand \bibinfo  [0]{\@secondoftwo}%
\providecommand \bibfield  [0]{\@secondoftwo}%
\providecommand \translation [1]{[#1]}%
\providecommand \BibitemOpen [0]{}%
\providecommand \bibitemStop [0]{}%
\providecommand \bibitemNoStop [0]{.\EOS\space}%
\providecommand \EOS [0]{\spacefactor3000\relax}%
\providecommand \BibitemShut  [1]{\csname bibitem#1\endcsname}%
\let\auto@bib@innerbib\@empty
\bibitem [{\citenamefont {Rigol}\ \emph {et~al.}(2008)\citenamefont {Rigol},
  \citenamefont {Dunjko},\ and\ \citenamefont
  {Olshanii}}]{rigol_thermalization}%
  \BibitemOpen
  \bibfield  {author} {\bibinfo {author} {\bibfnamefont {M.}~\bibnamefont
  {Rigol}}, \bibinfo {author} {\bibfnamefont {V.}~\bibnamefont {Dunjko}}, \
  and\ \bibinfo {author} {\bibfnamefont {M.}~\bibnamefont {Olshanii}},\ }\href
  {\doibase 10.1038/nature06838} {\bibfield  {journal} {\bibinfo  {journal}
  {Nature}\ }\textbf {\bibinfo {volume} {452}},\ \bibinfo {pages} {854}
  (\bibinfo {year} {2008})}\BibitemShut {NoStop}%
\bibitem [{\citenamefont {Deutsch}(1991)}]{deutsch}%
  \BibitemOpen
  \bibfield  {author} {\bibinfo {author} {\bibfnamefont {J.~M.}\ \bibnamefont
  {Deutsch}},\ }\href {\doibase 10.1103/PhysRevA.43.2046} {\bibfield  {journal}
  {\bibinfo  {journal} {Phys. Rev. A}\ }\textbf {\bibinfo {volume} {43}},\
  \bibinfo {pages} {2046} (\bibinfo {year} {1991})}\BibitemShut {NoStop}%
\bibitem [{\citenamefont {Srednicki}(1994)}]{srednicki}%
  \BibitemOpen
  \bibfield  {author} {\bibinfo {author} {\bibfnamefont {M.}~\bibnamefont
  {Srednicki}},\ }\href {\doibase 10.1103/PhysRevE.50.888} {\bibfield
  {journal} {\bibinfo  {journal} {Phys. Rev. E}\ }\textbf {\bibinfo {volume}
  {50}},\ \bibinfo {pages} {888} (\bibinfo {year} {1994})}\BibitemShut
  {NoStop}%
\bibitem [{\citenamefont {Basko}\ \emph {et~al.}(2006)\citenamefont {Basko},
  \citenamefont {Aleiner},\ and\ \citenamefont {Altshuler}}]{basko}%
  \BibitemOpen
  \bibfield  {author} {\bibinfo {author} {\bibfnamefont {D.}~\bibnamefont
  {Basko}}, \bibinfo {author} {\bibfnamefont {I.}~\bibnamefont {Aleiner}}, \
  and\ \bibinfo {author} {\bibfnamefont {B.}~\bibnamefont {Altshuler}},\ }\href
  {\doibase https://doi.org/10.1016/j.aop.2005.11.014} {\bibfield  {journal}
  {\bibinfo  {journal} {Annals of Physics}\ }\textbf {\bibinfo {volume}
  {321}},\ \bibinfo {pages} {1126 } (\bibinfo {year} {2006})}\BibitemShut
  {NoStop}%
\bibitem [{\citenamefont {Gornyi}\ \emph {et~al.}(2005)\citenamefont {Gornyi},
  \citenamefont {Mirlin},\ and\ \citenamefont
  {Polyakov}}]{gornyi_mirlin_polyakov}%
  \BibitemOpen
  \bibfield  {author} {\bibinfo {author} {\bibfnamefont {I.~V.}\ \bibnamefont
  {Gornyi}}, \bibinfo {author} {\bibfnamefont {A.~D.}\ \bibnamefont {Mirlin}},
  \ and\ \bibinfo {author} {\bibfnamefont {D.~G.}\ \bibnamefont {Polyakov}},\
  }\href {\doibase 10.1103/PhysRevLett.95.206603} {\bibfield  {journal}
  {\bibinfo  {journal} {Phys. Rev. Lett.}\ }\textbf {\bibinfo {volume} {95}},\
  \bibinfo {pages} {206603} (\bibinfo {year} {2005})}\BibitemShut {NoStop}%
\bibitem [{\citenamefont {Oganesyan}\ and\ \citenamefont
  {Huse}(2007)}]{oganesyan_huse}%
  \BibitemOpen
  \bibfield  {author} {\bibinfo {author} {\bibfnamefont {V.}~\bibnamefont
  {Oganesyan}}\ and\ \bibinfo {author} {\bibfnamefont {D.~A.}\ \bibnamefont
  {Huse}},\ }\href {\doibase 10.1103/PhysRevB.75.155111} {\bibfield  {journal}
  {\bibinfo  {journal} {Phys. Rev. B}\ }\textbf {\bibinfo {volume} {75}},\
  \bibinfo {pages} {155111} (\bibinfo {year} {2007})}\BibitemShut {NoStop}%
\bibitem [{\citenamefont {\ifmmode \check{Z}\else
  \v{Z}\fi{}nidari\ifmmode~\check{c}\else \v{c}\fi{}}\ \emph
  {et~al.}(2008)\citenamefont {\ifmmode \check{Z}\else
  \v{Z}\fi{}nidari\ifmmode~\check{c}\else \v{c}\fi{}}, \citenamefont {Prosen},\
  and\ \citenamefont {Prelov\ifmmode~\check{s}\else
  \v{s}\fi{}ek}}]{znidaric_prosen}%
  \BibitemOpen
  \bibfield  {author} {\bibinfo {author} {\bibfnamefont {M.}~\bibnamefont
  {\ifmmode \check{Z}\else \v{Z}\fi{}nidari\ifmmode~\check{c}\else
  \v{c}\fi{}}}, \bibinfo {author} {\bibfnamefont {T.~c.~v.}\ \bibnamefont
  {Prosen}}, \ and\ \bibinfo {author} {\bibfnamefont {P.}~\bibnamefont
  {Prelov\ifmmode~\check{s}\else \v{s}\fi{}ek}},\ }\href {\doibase
  10.1103/PhysRevB.77.064426} {\bibfield  {journal} {\bibinfo  {journal} {Phys.
  Rev. B}\ }\textbf {\bibinfo {volume} {77}},\ \bibinfo {pages} {064426}
  (\bibinfo {year} {2008})}\BibitemShut {NoStop}%
\bibitem [{\citenamefont {Pal}\ and\ \citenamefont {Huse}(2010)}]{pal_huse}%
  \BibitemOpen
  \bibfield  {author} {\bibinfo {author} {\bibfnamefont {A.}~\bibnamefont
  {Pal}}\ and\ \bibinfo {author} {\bibfnamefont {D.~A.}\ \bibnamefont {Huse}},\
  }\href {\doibase 10.1103/PhysRevB.82.174411} {\bibfield  {journal} {\bibinfo
  {journal} {Phys. Rev. B}\ }\textbf {\bibinfo {volume} {82}},\ \bibinfo
  {pages} {174411} (\bibinfo {year} {2010})}\BibitemShut {NoStop}%
\bibitem [{\citenamefont {Nandkishore}\ and\ \citenamefont
  {Huse}(2015)}]{nandkishore_huse}%
  \BibitemOpen
  \bibfield  {author} {\bibinfo {author} {\bibfnamefont {R.}~\bibnamefont
  {Nandkishore}}\ and\ \bibinfo {author} {\bibfnamefont {D.~A.}\ \bibnamefont
  {Huse}},\ }\href {\doibase 10.1146/annurev-conmatphys-031214-014726}
  {\bibfield  {journal} {\bibinfo  {journal} {Annual Review of Condensed Matter
  Physics}\ }\textbf {\bibinfo {volume} {6}},\ \bibinfo {pages} {15} (\bibinfo
  {year} {2015})}\BibitemShut {NoStop}%
\bibitem [{\citenamefont {Alet}\ and\ \citenamefont
  {Laflorencie}(2018)}]{alet_laflorencie}%
  \BibitemOpen
  \bibfield  {author} {\bibinfo {author} {\bibfnamefont {F.}~\bibnamefont
  {Alet}}\ and\ \bibinfo {author} {\bibfnamefont {N.}~\bibnamefont
  {Laflorencie}},\ }\href {\doibase https://doi.org/10.1016/j.crhy.2018.03.003}
  {\bibfield  {journal} {\bibinfo  {journal} {Comptes Rendus Physique}\
  }\textbf {\bibinfo {volume} {19}},\ \bibinfo {pages} {498 } (\bibinfo {year}
  {2018})},\ \bibinfo {note} {quantum simulation / Simulation
  quantique}\BibitemShut {NoStop}%
\bibitem [{\citenamefont {Abanin}\ \emph {et~al.}(2019)\citenamefont {Abanin},
  \citenamefont {Altman}, \citenamefont {Bloch},\ and\ \citenamefont
  {Serbyn}}]{abanin_altman_bloch_serbyn}%
  \BibitemOpen
  \bibfield  {author} {\bibinfo {author} {\bibfnamefont {D.~A.}\ \bibnamefont
  {Abanin}}, \bibinfo {author} {\bibfnamefont {E.}~\bibnamefont {Altman}},
  \bibinfo {author} {\bibfnamefont {I.}~\bibnamefont {Bloch}}, \ and\ \bibinfo
  {author} {\bibfnamefont {M.}~\bibnamefont {Serbyn}},\ }\href {\doibase
  10.1103/RevModPhys.91.021001} {\bibfield  {journal} {\bibinfo  {journal}
  {Rev. Mod. Phys.}\ }\textbf {\bibinfo {volume} {91}},\ \bibinfo {pages}
  {021001} (\bibinfo {year} {2019})}\BibitemShut {NoStop}%
\bibitem [{\citenamefont {Schreiber}\ \emph {et~al.}(2015)\citenamefont
  {Schreiber}, \citenamefont {Hodgman}, \citenamefont {Bordia}, \citenamefont
  {L{\"u}schen}, \citenamefont {Fischer}, \citenamefont {Vosk}, \citenamefont
  {Altman}, \citenamefont {Schneider},\ and\ \citenamefont
  {Bloch}}]{schreiber}%
  \BibitemOpen
  \bibfield  {author} {\bibinfo {author} {\bibfnamefont {M.}~\bibnamefont
  {Schreiber}}, \bibinfo {author} {\bibfnamefont {S.~S.}\ \bibnamefont
  {Hodgman}}, \bibinfo {author} {\bibfnamefont {P.}~\bibnamefont {Bordia}},
  \bibinfo {author} {\bibfnamefont {H.~P.}\ \bibnamefont {L{\"u}schen}},
  \bibinfo {author} {\bibfnamefont {M.~H.}\ \bibnamefont {Fischer}}, \bibinfo
  {author} {\bibfnamefont {R.}~\bibnamefont {Vosk}}, \bibinfo {author}
  {\bibfnamefont {E.}~\bibnamefont {Altman}}, \bibinfo {author} {\bibfnamefont
  {U.}~\bibnamefont {Schneider}}, \ and\ \bibinfo {author} {\bibfnamefont
  {I.}~\bibnamefont {Bloch}},\ }\href {\doibase 10.1126/science.aaa7432}
  {\bibfield  {journal} {\bibinfo  {journal} {Science}\ }\textbf {\bibinfo
  {volume} {349}},\ \bibinfo {pages} {842} (\bibinfo {year}
  {2015})}\BibitemShut {NoStop}%
\bibitem [{\citenamefont {Bordia}\ \emph {et~al.}(2016)\citenamefont {Bordia},
  \citenamefont {L\"uschen}, \citenamefont {Hodgman}, \citenamefont
  {Schreiber}, \citenamefont {Bloch},\ and\ \citenamefont
  {Schneider}}]{bordia_coupling}%
  \BibitemOpen
  \bibfield  {author} {\bibinfo {author} {\bibfnamefont {P.}~\bibnamefont
  {Bordia}}, \bibinfo {author} {\bibfnamefont {H.~P.}\ \bibnamefont
  {L\"uschen}}, \bibinfo {author} {\bibfnamefont {S.~S.}\ \bibnamefont
  {Hodgman}}, \bibinfo {author} {\bibfnamefont {M.}~\bibnamefont {Schreiber}},
  \bibinfo {author} {\bibfnamefont {I.}~\bibnamefont {Bloch}}, \ and\ \bibinfo
  {author} {\bibfnamefont {U.}~\bibnamefont {Schneider}},\ }\href {\doibase
  10.1103/PhysRevLett.116.140401} {\bibfield  {journal} {\bibinfo  {journal}
  {Phys. Rev. Lett.}\ }\textbf {\bibinfo {volume} {116}},\ \bibinfo {pages}
  {140401} (\bibinfo {year} {2016})}\BibitemShut {NoStop}%
\bibitem [{\citenamefont {Choi}\ \emph {et~al.}(2016)\citenamefont {Choi},
  \citenamefont {Hild}, \citenamefont {Zeiher}, \citenamefont {Schau{\ss}},
  \citenamefont {Rubio-Abadal}, \citenamefont {Yefsah}, \citenamefont
  {Khemani}, \citenamefont {Huse}, \citenamefont {Bloch},\ and\ \citenamefont
  {Gross}}]{choi}%
  \BibitemOpen
  \bibfield  {author} {\bibinfo {author} {\bibfnamefont {J.-y.}\ \bibnamefont
  {Choi}}, \bibinfo {author} {\bibfnamefont {S.}~\bibnamefont {Hild}}, \bibinfo
  {author} {\bibfnamefont {J.}~\bibnamefont {Zeiher}}, \bibinfo {author}
  {\bibfnamefont {P.}~\bibnamefont {Schau{\ss}}}, \bibinfo {author}
  {\bibfnamefont {A.}~\bibnamefont {Rubio-Abadal}}, \bibinfo {author}
  {\bibfnamefont {T.}~\bibnamefont {Yefsah}}, \bibinfo {author} {\bibfnamefont
  {V.}~\bibnamefont {Khemani}}, \bibinfo {author} {\bibfnamefont {D.~A.}\
  \bibnamefont {Huse}}, \bibinfo {author} {\bibfnamefont {I.}~\bibnamefont
  {Bloch}}, \ and\ \bibinfo {author} {\bibfnamefont {C.}~\bibnamefont
  {Gross}},\ }\href {\doibase 10.1126/science.aaf8834} {\bibfield  {journal}
  {\bibinfo  {journal} {Science}\ }\textbf {\bibinfo {volume} {352}},\ \bibinfo
  {pages} {1547} (\bibinfo {year} {2016})}\BibitemShut {NoStop}%
\bibitem [{\citenamefont {Vosk}\ \emph {et~al.}(2015)\citenamefont {Vosk},
  \citenamefont {Huse},\ and\ \citenamefont {Altman}}]{vosk_rg}%
  \BibitemOpen
  \bibfield  {author} {\bibinfo {author} {\bibfnamefont {R.}~\bibnamefont
  {Vosk}}, \bibinfo {author} {\bibfnamefont {D.~A.}\ \bibnamefont {Huse}}, \
  and\ \bibinfo {author} {\bibfnamefont {E.}~\bibnamefont {Altman}},\ }\href
  {\doibase 10.1103/PhysRevX.5.031032} {\bibfield  {journal} {\bibinfo
  {journal} {Phys. Rev. X}\ }\textbf {\bibinfo {volume} {5}},\ \bibinfo {pages}
  {031032} (\bibinfo {year} {2015})}\BibitemShut {NoStop}%
\bibitem [{\citenamefont {Potter}\ \emph {et~al.}(2015)\citenamefont {Potter},
  \citenamefont {Vasseur},\ and\ \citenamefont {Parameswaran}}]{potter_rg}%
  \BibitemOpen
  \bibfield  {author} {\bibinfo {author} {\bibfnamefont {A.~C.}\ \bibnamefont
  {Potter}}, \bibinfo {author} {\bibfnamefont {R.}~\bibnamefont {Vasseur}}, \
  and\ \bibinfo {author} {\bibfnamefont {S.~A.}\ \bibnamefont {Parameswaran}},\
  }\href {\doibase 10.1103/PhysRevX.5.031033} {\bibfield  {journal} {\bibinfo
  {journal} {Phys. Rev. X}\ }\textbf {\bibinfo {volume} {5}},\ \bibinfo {pages}
  {031033} (\bibinfo {year} {2015})}\BibitemShut {NoStop}%
\bibitem [{\citenamefont {Thiery}\ \emph {et~al.}(2018)\citenamefont {Thiery},
  \citenamefont {Huveneers}, \citenamefont {M\"uller},\ and\ \citenamefont
  {De~Roeck}}]{thiery_rg}%
  \BibitemOpen
  \bibfield  {author} {\bibinfo {author} {\bibfnamefont {T.}~\bibnamefont
  {Thiery}}, \bibinfo {author} {\bibfnamefont {F.~m.~c.}\ \bibnamefont
  {Huveneers}}, \bibinfo {author} {\bibfnamefont {M.}~\bibnamefont {M\"uller}},
  \ and\ \bibinfo {author} {\bibfnamefont {W.}~\bibnamefont {De~Roeck}},\
  }\href {\doibase 10.1103/PhysRevLett.121.140601} {\bibfield  {journal}
  {\bibinfo  {journal} {Phys. Rev. Lett.}\ }\textbf {\bibinfo {volume} {121}},\
  \bibinfo {pages} {140601} (\bibinfo {year} {2018})}\BibitemShut {NoStop}%
\bibitem [{\citenamefont {Imbrie}(2016)}]{imbrie}%
  \BibitemOpen
  \bibfield  {author} {\bibinfo {author} {\bibfnamefont {J.~Z.}\ \bibnamefont
  {Imbrie}},\ }\href {\doibase 10.1007/s10955-016-1508-x} {\bibfield  {journal}
  {\bibinfo  {journal} {Journal of Statistical Physics}\ }\textbf {\bibinfo
  {volume} {163}},\ \bibinfo {pages} {998} (\bibinfo {year}
  {2016})}\BibitemShut {NoStop}%
\bibitem [{\citenamefont {Luitz}\ \emph {et~al.}(2015)\citenamefont {Luitz},
  \citenamefont {Laflorencie},\ and\ \citenamefont
  {Alet}}]{luitz_laflorencie_alet}%
  \BibitemOpen
  \bibfield  {author} {\bibinfo {author} {\bibfnamefont {D.~J.}\ \bibnamefont
  {Luitz}}, \bibinfo {author} {\bibfnamefont {N.}~\bibnamefont {Laflorencie}},
  \ and\ \bibinfo {author} {\bibfnamefont {F.}~\bibnamefont {Alet}},\ }\href
  {\doibase 10.1103/PhysRevB.91.081103} {\bibfield  {journal} {\bibinfo
  {journal} {Phys. Rev. B}\ }\textbf {\bibinfo {volume} {91}},\ \bibinfo
  {pages} {081103} (\bibinfo {year} {2015})}\BibitemShut {NoStop}%
\bibitem [{\citenamefont {Bar~Lev}\ \emph {et~al.}(2015)\citenamefont
  {Bar~Lev}, \citenamefont {Cohen},\ and\ \citenamefont
  {Reichman}}]{barlev_cohen_reichman}%
  \BibitemOpen
  \bibfield  {author} {\bibinfo {author} {\bibfnamefont {Y.}~\bibnamefont
  {Bar~Lev}}, \bibinfo {author} {\bibfnamefont {G.}~\bibnamefont {Cohen}}, \
  and\ \bibinfo {author} {\bibfnamefont {D.~R.}\ \bibnamefont {Reichman}},\
  }\href {\doibase 10.1103/PhysRevLett.114.100601} {\bibfield  {journal}
  {\bibinfo  {journal} {Phys. Rev. Lett.}\ }\textbf {\bibinfo {volume} {114}},\
  \bibinfo {pages} {100601} (\bibinfo {year} {2015})}\BibitemShut {NoStop}%
\bibitem [{\citenamefont {Serbyn}\ \emph {et~al.}(2014)\citenamefont {Serbyn},
  \citenamefont {Papi\ifmmode~\acute{c}\else \'{c}\fi{}},\ and\ \citenamefont
  {Abanin}}]{serbyn_papic_abanin}%
  \BibitemOpen
  \bibfield  {author} {\bibinfo {author} {\bibfnamefont {M.}~\bibnamefont
  {Serbyn}}, \bibinfo {author} {\bibfnamefont {Z.}~\bibnamefont
  {Papi\ifmmode~\acute{c}\else \'{c}\fi{}}}, \ and\ \bibinfo {author}
  {\bibfnamefont {D.~A.}\ \bibnamefont {Abanin}},\ }\href {\doibase
  10.1103/PhysRevB.90.174302} {\bibfield  {journal} {\bibinfo  {journal} {Phys.
  Rev. B}\ }\textbf {\bibinfo {volume} {90}},\ \bibinfo {pages} {174302}
  (\bibinfo {year} {2014})}\BibitemShut {NoStop}%
\bibitem [{\citenamefont {Bardarson}\ \emph {et~al.}(2012)\citenamefont
  {Bardarson}, \citenamefont {Pollmann},\ and\ \citenamefont
  {Moore}}]{bardarson_jens_pollmann_moore}%
  \BibitemOpen
  \bibfield  {author} {\bibinfo {author} {\bibfnamefont {J.~H.}\ \bibnamefont
  {Bardarson}}, \bibinfo {author} {\bibfnamefont {F.}~\bibnamefont {Pollmann}},
  \ and\ \bibinfo {author} {\bibfnamefont {J.~E.}\ \bibnamefont {Moore}},\
  }\href {\doibase 10.1103/PhysRevLett.109.017202} {\bibfield  {journal}
  {\bibinfo  {journal} {Phys. Rev. Lett.}\ }\textbf {\bibinfo {volume} {109}},\
  \bibinfo {pages} {017202} (\bibinfo {year} {2012})}\BibitemShut {NoStop}%
\bibitem [{\citenamefont {Agarwal}\ \emph {et~al.}(2015)\citenamefont
  {Agarwal}, \citenamefont {Gopalakrishnan}, \citenamefont {Knap},
  \citenamefont {M\"uller},\ and\ \citenamefont
  {Demler}}]{agarwal_subdiffusion}%
  \BibitemOpen
  \bibfield  {author} {\bibinfo {author} {\bibfnamefont {K.}~\bibnamefont
  {Agarwal}}, \bibinfo {author} {\bibfnamefont {S.}~\bibnamefont
  {Gopalakrishnan}}, \bibinfo {author} {\bibfnamefont {M.}~\bibnamefont
  {Knap}}, \bibinfo {author} {\bibfnamefont {M.}~\bibnamefont {M\"uller}}, \
  and\ \bibinfo {author} {\bibfnamefont {E.}~\bibnamefont {Demler}},\ }\href
  {\doibase 10.1103/PhysRevLett.114.160401} {\bibfield  {journal} {\bibinfo
  {journal} {Phys. Rev. Lett.}\ }\textbf {\bibinfo {volume} {114}},\ \bibinfo
  {pages} {160401} (\bibinfo {year} {2015})}\BibitemShut {NoStop}%
\bibitem [{\citenamefont {\ifmmode \check{Z}\else
  \v{Z}\fi{}nidari\ifmmode~\check{c}\else \v{c}\fi{}}\ \emph
  {et~al.}(2016)\citenamefont {\ifmmode \check{Z}\else
  \v{Z}\fi{}nidari\ifmmode~\check{c}\else \v{c}\fi{}}, \citenamefont
  {Scardicchio},\ and\ \citenamefont {Varma}}]{znidaric_subdiffusion}%
  \BibitemOpen
  \bibfield  {author} {\bibinfo {author} {\bibfnamefont {M.}~\bibnamefont
  {\ifmmode \check{Z}\else \v{Z}\fi{}nidari\ifmmode~\check{c}\else
  \v{c}\fi{}}}, \bibinfo {author} {\bibfnamefont {A.}~\bibnamefont
  {Scardicchio}}, \ and\ \bibinfo {author} {\bibfnamefont {V.~K.}\ \bibnamefont
  {Varma}},\ }\href {\doibase 10.1103/PhysRevLett.117.040601} {\bibfield
  {journal} {\bibinfo  {journal} {Phys. Rev. Lett.}\ }\textbf {\bibinfo
  {volume} {117}},\ \bibinfo {pages} {040601} (\bibinfo {year}
  {2016})}\BibitemShut {NoStop}%
\bibitem [{\citenamefont {Khait}\ \emph {et~al.}(2016)\citenamefont {Khait},
  \citenamefont {Gazit}, \citenamefont {Yao},\ and\ \citenamefont
  {Auerbach}}]{khait_subdiffusion}%
  \BibitemOpen
  \bibfield  {author} {\bibinfo {author} {\bibfnamefont {I.}~\bibnamefont
  {Khait}}, \bibinfo {author} {\bibfnamefont {S.}~\bibnamefont {Gazit}},
  \bibinfo {author} {\bibfnamefont {N.~Y.}\ \bibnamefont {Yao}}, \ and\
  \bibinfo {author} {\bibfnamefont {A.}~\bibnamefont {Auerbach}},\ }\href
  {\doibase 10.1103/PhysRevB.93.224205} {\bibfield  {journal} {\bibinfo
  {journal} {Phys. Rev. B}\ }\textbf {\bibinfo {volume} {93}},\ \bibinfo
  {pages} {224205} (\bibinfo {year} {2016})}\BibitemShut {NoStop}%
\bibitem [{\citenamefont {Luitz}\ and\ \citenamefont
  {Bar~Lev}(2017{\natexlab{a}})}]{luitz_subdiffusion}%
  \BibitemOpen
  \bibfield  {author} {\bibinfo {author} {\bibfnamefont {D.~J.}\ \bibnamefont
  {Luitz}}\ and\ \bibinfo {author} {\bibfnamefont {Y.}~\bibnamefont
  {Bar~Lev}},\ }\href {\doibase 10.1103/PhysRevB.96.020406} {\bibfield
  {journal} {\bibinfo  {journal} {Phys. Rev. B}\ }\textbf {\bibinfo {volume}
  {96}},\ \bibinfo {pages} {020406} (\bibinfo {year}
  {2017}{\natexlab{a}})}\BibitemShut {NoStop}%
\bibitem [{\citenamefont {Serbyn}\ \emph {et~al.}(2017)\citenamefont {Serbyn},
  \citenamefont {Papi\ifmmode~\acute{c}\else \'{c}\fi{}},\ and\ \citenamefont
  {Abanin}}]{serbyn_thouless}%
  \BibitemOpen
  \bibfield  {author} {\bibinfo {author} {\bibfnamefont {M.}~\bibnamefont
  {Serbyn}}, \bibinfo {author} {\bibfnamefont {Z.}~\bibnamefont
  {Papi\ifmmode~\acute{c}\else \'{c}\fi{}}}, \ and\ \bibinfo {author}
  {\bibfnamefont {D.~A.}\ \bibnamefont {Abanin}},\ }\href {\doibase
  10.1103/PhysRevB.96.104201} {\bibfield  {journal} {\bibinfo  {journal} {Phys.
  Rev. B}\ }\textbf {\bibinfo {volume} {96}},\ \bibinfo {pages} {104201}
  (\bibinfo {year} {2017})}\BibitemShut {NoStop}%
\bibitem [{\citenamefont {Lezama}\ \emph {et~al.}(2019)\citenamefont {Lezama},
  \citenamefont {Bera},\ and\ \citenamefont
  {Bardarson}}]{lezama_slow_dynamics}%
  \BibitemOpen
  \bibfield  {author} {\bibinfo {author} {\bibfnamefont {T.~L.~M.}\
  \bibnamefont {Lezama}}, \bibinfo {author} {\bibfnamefont {S.}~\bibnamefont
  {Bera}}, \ and\ \bibinfo {author} {\bibfnamefont {J.~H.}\ \bibnamefont
  {Bardarson}},\ }\href {\doibase 10.1103/PhysRevB.99.161106} {\bibfield
  {journal} {\bibinfo  {journal} {Phys. Rev. B}\ }\textbf {\bibinfo {volume}
  {99}},\ \bibinfo {pages} {161106} (\bibinfo {year} {2019})}\BibitemShut
  {NoStop}%
\bibitem [{\citenamefont {Luitz}\ and\ \citenamefont
  {Bar~Lev}(2017{\natexlab{b}})}]{luitz_barlev_review}%
  \BibitemOpen
  \bibfield  {author} {\bibinfo {author} {\bibfnamefont {D.~J.}\ \bibnamefont
  {Luitz}}\ and\ \bibinfo {author} {\bibfnamefont {Y.}~\bibnamefont
  {Bar~Lev}},\ }\href {\doibase 10.1002/andp.201600350} {\bibfield  {journal}
  {\bibinfo  {journal} {Annalen der Physik}\ }\textbf {\bibinfo {volume}
  {529}},\ \bibinfo {pages} {1600350} (\bibinfo {year}
  {2017}{\natexlab{b}})}\BibitemShut {NoStop}%
\bibitem [{\citenamefont {Agarwal}\ \emph {et~al.}(2017)\citenamefont
  {Agarwal}, \citenamefont {Altman}, \citenamefont {Demler}, \citenamefont
  {Gopalakrishnan}, \citenamefont {Huse},\ and\ \citenamefont
  {Knap}}]{agarwal_subdiffusion_review}%
  \BibitemOpen
  \bibfield  {author} {\bibinfo {author} {\bibfnamefont {K.}~\bibnamefont
  {Agarwal}}, \bibinfo {author} {\bibfnamefont {E.}~\bibnamefont {Altman}},
  \bibinfo {author} {\bibfnamefont {E.}~\bibnamefont {Demler}}, \bibinfo
  {author} {\bibfnamefont {S.}~\bibnamefont {Gopalakrishnan}}, \bibinfo
  {author} {\bibfnamefont {D.~A.}\ \bibnamefont {Huse}}, \ and\ \bibinfo
  {author} {\bibfnamefont {M.}~\bibnamefont {Knap}},\ }\href {\doibase
  10.1002/andp.201600326} {\bibfield  {journal} {\bibinfo  {journal} {Annalen
  der Physik}\ }\textbf {\bibinfo {volume} {529}},\ \bibinfo {pages} {1600326}
  (\bibinfo {year} {2017})}\BibitemShut {NoStop}%
\bibitem [{\citenamefont {Goremykina}\ \emph {et~al.}(2019)\citenamefont
  {Goremykina}, \citenamefont {Vasseur},\ and\ \citenamefont
  {Serbyn}}]{goremykina_rg}%
  \BibitemOpen
  \bibfield  {author} {\bibinfo {author} {\bibfnamefont {A.}~\bibnamefont
  {Goremykina}}, \bibinfo {author} {\bibfnamefont {R.}~\bibnamefont {Vasseur}},
  \ and\ \bibinfo {author} {\bibfnamefont {M.}~\bibnamefont {Serbyn}},\ }\href
  {\doibase 10.1103/PhysRevLett.122.040601} {\bibfield  {journal} {\bibinfo
  {journal} {Phys. Rev. Lett.}\ }\textbf {\bibinfo {volume} {122}},\ \bibinfo
  {pages} {040601} (\bibinfo {year} {2019})}\BibitemShut {NoStop}%
\bibitem [{\citenamefont {Morningstar}\ and\ \citenamefont
  {Huse}(2019)}]{morningstar_rg}%
  \BibitemOpen
  \bibfield  {author} {\bibinfo {author} {\bibfnamefont {A.}~\bibnamefont
  {Morningstar}}\ and\ \bibinfo {author} {\bibfnamefont {D.~A.}\ \bibnamefont
  {Huse}},\ }\href {\doibase 10.1103/PhysRevB.99.224205} {\bibfield  {journal}
  {\bibinfo  {journal} {Phys. Rev. B}\ }\textbf {\bibinfo {volume} {99}},\
  \bibinfo {pages} {224205} (\bibinfo {year} {2019})}\BibitemShut {NoStop}%
\bibitem [{\citenamefont {Schir{\'o}}\ and\ \citenamefont
  {Tarzia}(2019)}]{schiro_tarzia}%
  \BibitemOpen
  \bibfield  {author} {\bibinfo {author} {\bibfnamefont {M.}~\bibnamefont
  {Schir{\'o}}}\ and\ \bibinfo {author} {\bibfnamefont {M.}~\bibnamefont
  {Tarzia}},\ }\href@noop {} {\bibfield  {journal} {\bibinfo  {journal} {arXiv
  preprint arXiv:1909.07160}\ } (\bibinfo {year} {2019})}\BibitemShut {NoStop}%
\bibitem [{\citenamefont {Zhang}\ \emph {et~al.}(2016)\citenamefont {Zhang},
  \citenamefont {Khemani},\ and\ \citenamefont {Huse}}]{zhang_floquet_mbl}%
  \BibitemOpen
  \bibfield  {author} {\bibinfo {author} {\bibfnamefont {L.}~\bibnamefont
  {Zhang}}, \bibinfo {author} {\bibfnamefont {V.}~\bibnamefont {Khemani}}, \
  and\ \bibinfo {author} {\bibfnamefont {D.~A.}\ \bibnamefont {Huse}},\ }\href
  {\doibase 10.1103/PhysRevB.94.224202} {\bibfield  {journal} {\bibinfo
  {journal} {Phys. Rev. B}\ }\textbf {\bibinfo {volume} {94}},\ \bibinfo
  {pages} {224202} (\bibinfo {year} {2016})}\BibitemShut {NoStop}%
\bibitem [{\citenamefont {Abanin}\ \emph {et~al.}(2016)\citenamefont {Abanin},
  \citenamefont {Roeck},\ and\ \citenamefont
  {Huveneers}}]{abanin_theory_floquet}%
  \BibitemOpen
  \bibfield  {author} {\bibinfo {author} {\bibfnamefont {D.~A.}\ \bibnamefont
  {Abanin}}, \bibinfo {author} {\bibfnamefont {W.~D.}\ \bibnamefont {Roeck}}, \
  and\ \bibinfo {author} {\bibfnamefont {F.}~\bibnamefont {Huveneers}},\ }\href
  {\doibase https://doi.org/10.1016/j.aop.2016.03.010} {\bibfield  {journal}
  {\bibinfo  {journal} {Annals of Physics}\ }\textbf {\bibinfo {volume}
  {372}},\ \bibinfo {pages} {1 } (\bibinfo {year} {2016})}\BibitemShut
  {NoStop}%
\bibitem [{\citenamefont {Yao}\ \emph {et~al.}(2017)\citenamefont {Yao},
  \citenamefont {Potter}, \citenamefont {Potirniche},\ and\ \citenamefont
  {Vishwanath}}]{yao_time_crystal}%
  \BibitemOpen
  \bibfield  {author} {\bibinfo {author} {\bibfnamefont {N.~Y.}\ \bibnamefont
  {Yao}}, \bibinfo {author} {\bibfnamefont {A.~C.}\ \bibnamefont {Potter}},
  \bibinfo {author} {\bibfnamefont {I.-D.}\ \bibnamefont {Potirniche}}, \ and\
  \bibinfo {author} {\bibfnamefont {A.}~\bibnamefont {Vishwanath}},\ }\href
  {\doibase 10.1103/PhysRevLett.118.030401} {\bibfield  {journal} {\bibinfo
  {journal} {Phys. Rev. Lett.}\ }\textbf {\bibinfo {volume} {118}},\ \bibinfo
  {pages} {030401} (\bibinfo {year} {2017})}\BibitemShut {NoStop}%
\bibitem [{\citenamefont {Ponte}\ \emph {et~al.}(2015)\citenamefont {Ponte},
  \citenamefont {Papi\ifmmode~\acute{c}\else \'{c}\fi{}}, \citenamefont
  {Huveneers},\ and\ \citenamefont {Abanin}}]{ponte_noncircuit_floquet}%
  \BibitemOpen
  \bibfield  {author} {\bibinfo {author} {\bibfnamefont {P.}~\bibnamefont
  {Ponte}}, \bibinfo {author} {\bibfnamefont {Z.}~\bibnamefont
  {Papi\ifmmode~\acute{c}\else \'{c}\fi{}}}, \bibinfo {author} {\bibfnamefont
  {F.~m.~c.}\ \bibnamefont {Huveneers}}, \ and\ \bibinfo {author}
  {\bibfnamefont {D.~A.}\ \bibnamefont {Abanin}},\ }\href {\doibase
  10.1103/PhysRevLett.114.140401} {\bibfield  {journal} {\bibinfo  {journal}
  {Phys. Rev. Lett.}\ }\textbf {\bibinfo {volume} {114}},\ \bibinfo {pages}
  {140401} (\bibinfo {year} {2015})}\BibitemShut {NoStop}%
\bibitem [{\citenamefont {Lazarides}\ \emph {et~al.}(2015)\citenamefont
  {Lazarides}, \citenamefont {Das},\ and\ \citenamefont
  {Moessner}}]{lazarides_noncircuit_floquet}%
  \BibitemOpen
  \bibfield  {author} {\bibinfo {author} {\bibfnamefont {A.}~\bibnamefont
  {Lazarides}}, \bibinfo {author} {\bibfnamefont {A.}~\bibnamefont {Das}}, \
  and\ \bibinfo {author} {\bibfnamefont {R.}~\bibnamefont {Moessner}},\ }\href
  {\doibase 10.1103/PhysRevLett.115.030402} {\bibfield  {journal} {\bibinfo
  {journal} {Phys. Rev. Lett.}\ }\textbf {\bibinfo {volume} {115}},\ \bibinfo
  {pages} {030402} (\bibinfo {year} {2015})}\BibitemShut {NoStop}%
\bibitem [{\citenamefont {Bordia}\ \emph {et~al.}(2017)\citenamefont {Bordia},
  \citenamefont {L{\"u}schen}, \citenamefont {Schneider}, \citenamefont
  {Knap},\ and\ \citenamefont {Bloch}}]{bordia_floquet_experiment}%
  \BibitemOpen
  \bibfield  {author} {\bibinfo {author} {\bibfnamefont {P.}~\bibnamefont
  {Bordia}}, \bibinfo {author} {\bibfnamefont {H.}~\bibnamefont {L{\"u}schen}},
  \bibinfo {author} {\bibfnamefont {U.}~\bibnamefont {Schneider}}, \bibinfo
  {author} {\bibfnamefont {M.}~\bibnamefont {Knap}}, \ and\ \bibinfo {author}
  {\bibfnamefont {I.}~\bibnamefont {Bloch}},\ }\href {\doibase
  10.1038/nphys4020} {\bibfield  {journal} {\bibinfo  {journal} {Nature
  Physics}\ }\textbf {\bibinfo {volume} {13}},\ \bibinfo {pages} {460}
  (\bibinfo {year} {2017})}\BibitemShut {NoStop}%
\bibitem [{\citenamefont {Else}\ \emph {et~al.}(2016)\citenamefont {Else},
  \citenamefont {Bauer},\ and\ \citenamefont {Nayak}}]{else_time_crystal}%
  \BibitemOpen
  \bibfield  {author} {\bibinfo {author} {\bibfnamefont {D.~V.}\ \bibnamefont
  {Else}}, \bibinfo {author} {\bibfnamefont {B.}~\bibnamefont {Bauer}}, \ and\
  \bibinfo {author} {\bibfnamefont {C.}~\bibnamefont {Nayak}},\ }\href
  {\doibase 10.1103/PhysRevLett.117.090402} {\bibfield  {journal} {\bibinfo
  {journal} {Phys. Rev. Lett.}\ }\textbf {\bibinfo {volume} {117}},\ \bibinfo
  {pages} {090402} (\bibinfo {year} {2016})}\BibitemShut {NoStop}%
\bibitem [{\citenamefont {Khemani}\ \emph {et~al.}(2016)\citenamefont
  {Khemani}, \citenamefont {Lazarides}, \citenamefont {Moessner},\ and\
  \citenamefont {Sondhi}}]{khemani_time_crystal}%
  \BibitemOpen
  \bibfield  {author} {\bibinfo {author} {\bibfnamefont {V.}~\bibnamefont
  {Khemani}}, \bibinfo {author} {\bibfnamefont {A.}~\bibnamefont {Lazarides}},
  \bibinfo {author} {\bibfnamefont {R.}~\bibnamefont {Moessner}}, \ and\
  \bibinfo {author} {\bibfnamefont {S.~L.}\ \bibnamefont {Sondhi}},\ }\href
  {\doibase 10.1103/PhysRevLett.116.250401} {\bibfield  {journal} {\bibinfo
  {journal} {Phys. Rev. Lett.}\ }\textbf {\bibinfo {volume} {116}},\ \bibinfo
  {pages} {250401} (\bibinfo {year} {2016})}\BibitemShut {NoStop}%
\bibitem [{\citenamefont {\ifmmode \check{Z}\else
  \v{Z}\fi{}nidari\ifmmode~\check{c}\else
  \v{c}\fi{}}(2007)}]{znidaric_entanglement}%
  \BibitemOpen
  \bibfield  {author} {\bibinfo {author} {\bibfnamefont {M.}~\bibnamefont
  {\ifmmode \check{Z}\else \v{Z}\fi{}nidari\ifmmode~\check{c}\else
  \v{c}\fi{}}},\ }\href {\doibase 10.1103/PhysRevA.76.012318} {\bibfield
  {journal} {\bibinfo  {journal} {Phys. Rev. A}\ }\textbf {\bibinfo {volume}
  {76}},\ \bibinfo {pages} {012318} (\bibinfo {year} {2007})}\BibitemShut
  {NoStop}%
\bibitem [{\citenamefont {Oliveira}\ \emph {et~al.}(2007)\citenamefont
  {Oliveira}, \citenamefont {Dahlsten},\ and\ \citenamefont
  {Plenio}}]{oliveira_entanglement}%
  \BibitemOpen
  \bibfield  {author} {\bibinfo {author} {\bibfnamefont {R.}~\bibnamefont
  {Oliveira}}, \bibinfo {author} {\bibfnamefont {O.~C.~O.}\ \bibnamefont
  {Dahlsten}}, \ and\ \bibinfo {author} {\bibfnamefont {M.~B.}\ \bibnamefont
  {Plenio}},\ }\href {\doibase 10.1103/PhysRevLett.98.130502} {\bibfield
  {journal} {\bibinfo  {journal} {Phys. Rev. Lett.}\ }\textbf {\bibinfo
  {volume} {98}},\ \bibinfo {pages} {130502} (\bibinfo {year}
  {2007})}\BibitemShut {NoStop}%
\bibitem [{\citenamefont {G\"utschow}\ \emph {et~al.}(2010)\citenamefont
  {G\"utschow}, \citenamefont {Uphoff}, \citenamefont {Werner},\ and\
  \citenamefont {Zimbor\'as}}]{gutschow_entanglement_clifford}%
  \BibitemOpen
  \bibfield  {author} {\bibinfo {author} {\bibfnamefont {J.}~\bibnamefont
  {G\"utschow}}, \bibinfo {author} {\bibfnamefont {S.}~\bibnamefont {Uphoff}},
  \bibinfo {author} {\bibfnamefont {R.~F.}\ \bibnamefont {Werner}}, \ and\
  \bibinfo {author} {\bibfnamefont {Z.}~\bibnamefont {Zimbor\'as}},\ }\href
  {\doibase 10.1063/1.3278513} {\bibfield  {journal} {\bibinfo  {journal}
  {Journal of Mathematical Physics}\ }\textbf {\bibinfo {volume} {51}},\
  \bibinfo {pages} {015203} (\bibinfo {year} {2010})}\BibitemShut {NoStop}%
\bibitem [{\citenamefont {Nahum}\ \emph {et~al.}(2017)\citenamefont {Nahum},
  \citenamefont {Ruhman}, \citenamefont {Vijay},\ and\ \citenamefont
  {Haah}}]{nahum_entanglement_growth}%
  \BibitemOpen
  \bibfield  {author} {\bibinfo {author} {\bibfnamefont {A.}~\bibnamefont
  {Nahum}}, \bibinfo {author} {\bibfnamefont {J.}~\bibnamefont {Ruhman}},
  \bibinfo {author} {\bibfnamefont {S.}~\bibnamefont {Vijay}}, \ and\ \bibinfo
  {author} {\bibfnamefont {J.}~\bibnamefont {Haah}},\ }\href {\doibase
  10.1103/PhysRevX.7.031016} {\bibfield  {journal} {\bibinfo  {journal} {Phys.
  Rev. X}\ }\textbf {\bibinfo {volume} {7}},\ \bibinfo {pages} {031016}
  (\bibinfo {year} {2017})}\BibitemShut {NoStop}%
\bibitem [{\citenamefont {Nahum}\ \emph
  {et~al.}(2018{\natexlab{a}})\citenamefont {Nahum}, \citenamefont {Ruhman},\
  and\ \citenamefont {Huse}}]{nahum_entanglement_quenched}%
  \BibitemOpen
  \bibfield  {author} {\bibinfo {author} {\bibfnamefont {A.}~\bibnamefont
  {Nahum}}, \bibinfo {author} {\bibfnamefont {J.}~\bibnamefont {Ruhman}}, \
  and\ \bibinfo {author} {\bibfnamefont {D.~A.}\ \bibnamefont {Huse}},\ }\href
  {\doibase 10.1103/PhysRevB.98.035118} {\bibfield  {journal} {\bibinfo
  {journal} {Phys. Rev. B}\ }\textbf {\bibinfo {volume} {98}},\ \bibinfo
  {pages} {035118} (\bibinfo {year} {2018}{\natexlab{a}})}\BibitemShut
  {NoStop}%
\bibitem [{\citenamefont {von Keyserlingk}\ \emph {et~al.}(2018)\citenamefont
  {von Keyserlingk}, \citenamefont {Rakovszky}, \citenamefont {Pollmann},\ and\
  \citenamefont {Sondhi}}]{vonkeyserlingk_entanglement_OTOC}%
  \BibitemOpen
  \bibfield  {author} {\bibinfo {author} {\bibfnamefont {C.~W.}\ \bibnamefont
  {von Keyserlingk}}, \bibinfo {author} {\bibfnamefont {T.}~\bibnamefont
  {Rakovszky}}, \bibinfo {author} {\bibfnamefont {F.}~\bibnamefont {Pollmann}},
  \ and\ \bibinfo {author} {\bibfnamefont {S.~L.}\ \bibnamefont {Sondhi}},\
  }\href {\doibase 10.1103/PhysRevX.8.021013} {\bibfield  {journal} {\bibinfo
  {journal} {Phys. Rev. X}\ }\textbf {\bibinfo {volume} {8}},\ \bibinfo {pages}
  {021013} (\bibinfo {year} {2018})}\BibitemShut {NoStop}%
\bibitem [{\citenamefont {Jonay}\ \emph {et~al.}(2018)\citenamefont {Jonay},
  \citenamefont {Huse},\ and\ \citenamefont
  {Nahum}}]{nahum_operator_entanglement}%
  \BibitemOpen
  \bibfield  {author} {\bibinfo {author} {\bibfnamefont {C.}~\bibnamefont
  {Jonay}}, \bibinfo {author} {\bibfnamefont {D.~A.}\ \bibnamefont {Huse}}, \
  and\ \bibinfo {author} {\bibfnamefont {A.}~\bibnamefont {Nahum}},\
  }\href@noop {} {\bibfield  {journal} {\bibinfo  {journal} {arXiv preprint
  arXiv:1803.00089}\ } (\bibinfo {year} {2018})}\BibitemShut {NoStop}%
\bibitem [{\citenamefont {Rakovszky}\ \emph
  {et~al.}(2019{\natexlab{a}})\citenamefont {Rakovszky}, \citenamefont
  {Pollmann},\ and\ \citenamefont {von
  Keyserlingk}}]{rakovszky_subballistic_entanglement}%
  \BibitemOpen
  \bibfield  {author} {\bibinfo {author} {\bibfnamefont {T.}~\bibnamefont
  {Rakovszky}}, \bibinfo {author} {\bibfnamefont {F.}~\bibnamefont {Pollmann}},
  \ and\ \bibinfo {author} {\bibfnamefont {C.~W.}\ \bibnamefont {von
  Keyserlingk}},\ }\href {\doibase 10.1103/PhysRevLett.122.250602} {\bibfield
  {journal} {\bibinfo  {journal} {Phys. Rev. Lett.}\ }\textbf {\bibinfo
  {volume} {122}},\ \bibinfo {pages} {250602} (\bibinfo {year}
  {2019}{\natexlab{a}})}\BibitemShut {NoStop}%
\bibitem [{\citenamefont {Huang}(2019)}]{huang_subballistic_entanglement}%
  \BibitemOpen
  \bibfield  {author} {\bibinfo {author} {\bibfnamefont {Y.}~\bibnamefont
  {Huang}},\ }\href@noop {} {\bibfield  {journal} {\bibinfo  {journal} {arXiv
  preprint arXiv:1902.00977}\ } (\bibinfo {year} {2019})}\BibitemShut {NoStop}%
\bibitem [{\citenamefont {Rakovszky}\ \emph
  {et~al.}(2019{\natexlab{b}})\citenamefont {Rakovszky}, \citenamefont {von
  Keyserlingk},\ and\ \citenamefont
  {Pollmann}}]{rakovszky_inhomogenous_quench}%
  \BibitemOpen
  \bibfield  {author} {\bibinfo {author} {\bibfnamefont {T.}~\bibnamefont
  {Rakovszky}}, \bibinfo {author} {\bibfnamefont {C.~W.}\ \bibnamefont {von
  Keyserlingk}}, \ and\ \bibinfo {author} {\bibfnamefont {F.}~\bibnamefont
  {Pollmann}},\ }\href {\doibase 10.1103/PhysRevB.100.125139} {\bibfield
  {journal} {\bibinfo  {journal} {Phys. Rev. B}\ }\textbf {\bibinfo {volume}
  {100}},\ \bibinfo {pages} {125139} (\bibinfo {year}
  {2019}{\natexlab{b}})}\BibitemShut {NoStop}%
\bibitem [{\citenamefont {Nahum}\ \emph
  {et~al.}(2018{\natexlab{b}})\citenamefont {Nahum}, \citenamefont {Vijay},\
  and\ \citenamefont {Haah}}]{nahum_operator_spreading}%
  \BibitemOpen
  \bibfield  {author} {\bibinfo {author} {\bibfnamefont {A.}~\bibnamefont
  {Nahum}}, \bibinfo {author} {\bibfnamefont {S.}~\bibnamefont {Vijay}}, \ and\
  \bibinfo {author} {\bibfnamefont {J.}~\bibnamefont {Haah}},\ }\href {\doibase
  10.1103/PhysRevX.8.021014} {\bibfield  {journal} {\bibinfo  {journal} {Phys.
  Rev. X}\ }\textbf {\bibinfo {volume} {8}},\ \bibinfo {pages} {021014}
  (\bibinfo {year} {2018}{\natexlab{b}})}\BibitemShut {NoStop}%
\bibitem [{\citenamefont {Roberts}\ and\ \citenamefont
  {Yoshida}(2017)}]{roberts_frame_potential_otoc}%
  \BibitemOpen
  \bibfield  {author} {\bibinfo {author} {\bibfnamefont {D.~A.}\ \bibnamefont
  {Roberts}}\ and\ \bibinfo {author} {\bibfnamefont {B.}~\bibnamefont
  {Yoshida}},\ }\href {\doibase 10.1007/JHEP04(2017)121} {\bibfield  {journal}
  {\bibinfo  {journal} {Journal of High Energy Physics}\ }\textbf {\bibinfo
  {volume} {2017}},\ \bibinfo {pages} {121} (\bibinfo {year}
  {2017})}\BibitemShut {NoStop}%
\bibitem [{\citenamefont {Rakovszky}\ \emph {et~al.}(2018)\citenamefont
  {Rakovszky}, \citenamefont {Pollmann},\ and\ \citenamefont {von
  Keyserlingk}}]{rakovszky_otoc_conserved}%
  \BibitemOpen
  \bibfield  {author} {\bibinfo {author} {\bibfnamefont {T.}~\bibnamefont
  {Rakovszky}}, \bibinfo {author} {\bibfnamefont {F.}~\bibnamefont {Pollmann}},
  \ and\ \bibinfo {author} {\bibfnamefont {C.~W.}\ \bibnamefont {von
  Keyserlingk}},\ }\href {\doibase 10.1103/PhysRevX.8.031058} {\bibfield
  {journal} {\bibinfo  {journal} {Phys. Rev. X}\ }\textbf {\bibinfo {volume}
  {8}},\ \bibinfo {pages} {031058} (\bibinfo {year} {2018})}\BibitemShut
  {NoStop}%
\bibitem [{\citenamefont {Chan}\ \emph {et~al.}(2018)\citenamefont {Chan},
  \citenamefont {De~Luca},\ and\ \citenamefont {Chalker}}]{chan_dephasing}%
  \BibitemOpen
  \bibfield  {author} {\bibinfo {author} {\bibfnamefont {A.}~\bibnamefont
  {Chan}}, \bibinfo {author} {\bibfnamefont {A.}~\bibnamefont {De~Luca}}, \
  and\ \bibinfo {author} {\bibfnamefont {J.~T.}\ \bibnamefont {Chalker}},\
  }\href {\doibase 10.1103/PhysRevLett.121.060601} {\bibfield  {journal}
  {\bibinfo  {journal} {Phys. Rev. Lett.}\ }\textbf {\bibinfo {volume} {121}},\
  \bibinfo {pages} {060601} (\bibinfo {year} {2018})}\BibitemShut {NoStop}%
\bibitem [{\citenamefont {Atas}\ \emph {et~al.}(2013)\citenamefont {Atas},
  \citenamefont {Bogomolny}, \citenamefont {Giraud},\ and\ \citenamefont
  {Roux}}]{atas_rgaps}%
  \BibitemOpen
  \bibfield  {author} {\bibinfo {author} {\bibfnamefont {Y.~Y.}\ \bibnamefont
  {Atas}}, \bibinfo {author} {\bibfnamefont {E.}~\bibnamefont {Bogomolny}},
  \bibinfo {author} {\bibfnamefont {O.}~\bibnamefont {Giraud}}, \ and\ \bibinfo
  {author} {\bibfnamefont {G.}~\bibnamefont {Roux}},\ }\href {\doibase
  10.1103/PhysRevLett.110.084101} {\bibfield  {journal} {\bibinfo  {journal}
  {Phys. Rev. Lett.}\ }\textbf {\bibinfo {volume} {110}},\ \bibinfo {pages}
  {084101} (\bibinfo {year} {2013})}\BibitemShut {NoStop}%
\bibitem [{\citenamefont {D'Alessio}\ and\ \citenamefont
  {Rigol}(2014)}]{dalessio_rgaps_floquet}%
  \BibitemOpen
  \bibfield  {author} {\bibinfo {author} {\bibfnamefont {L.}~\bibnamefont
  {D'Alessio}}\ and\ \bibinfo {author} {\bibfnamefont {M.}~\bibnamefont
  {Rigol}},\ }\href {\doibase 10.1103/PhysRevX.4.041048} {\bibfield  {journal}
  {\bibinfo  {journal} {Phys. Rev. X}\ }\textbf {\bibinfo {volume} {4}},\
  \bibinfo {pages} {041048} (\bibinfo {year} {2014})}\BibitemShut {NoStop}%
\bibitem [{\citenamefont {Kj\"all}\ \emph {et~al.}(2014)\citenamefont
  {Kj\"all}, \citenamefont {Bardarson},\ and\ \citenamefont
  {Pollmann}}]{kjall_entropy}%
  \BibitemOpen
  \bibfield  {author} {\bibinfo {author} {\bibfnamefont {J.~A.}\ \bibnamefont
  {Kj\"all}}, \bibinfo {author} {\bibfnamefont {J.~H.}\ \bibnamefont
  {Bardarson}}, \ and\ \bibinfo {author} {\bibfnamefont {F.}~\bibnamefont
  {Pollmann}},\ }\href {\doibase 10.1103/PhysRevLett.113.107204} {\bibfield
  {journal} {\bibinfo  {journal} {Phys. Rev. Lett.}\ }\textbf {\bibinfo
  {volume} {113}},\ \bibinfo {pages} {107204} (\bibinfo {year}
  {2014})}\BibitemShut {NoStop}%
\bibitem [{\citenamefont {Page}(1993)}]{page}%
  \BibitemOpen
  \bibfield  {author} {\bibinfo {author} {\bibfnamefont {D.~N.}\ \bibnamefont
  {Page}},\ }\href {\doibase 10.1103/PhysRevLett.71.1291} {\bibfield  {journal}
  {\bibinfo  {journal} {Phys. Rev. Lett.}\ }\textbf {\bibinfo {volume} {71}},\
  \bibinfo {pages} {1291} (\bibinfo {year} {1993})}\BibitemShut {NoStop}%
\bibitem [{\citenamefont {Bauer}\ and\ \citenamefont
  {Nayak}(2013)}]{bauer_nayak}%
  \BibitemOpen
  \bibfield  {author} {\bibinfo {author} {\bibfnamefont {B.}~\bibnamefont
  {Bauer}}\ and\ \bibinfo {author} {\bibfnamefont {C.}~\bibnamefont {Nayak}},\
  }\href {\doibase 10.1088/1742-5468/2013/09/p09005} {\bibfield  {journal}
  {\bibinfo  {journal} {Journal of Statistical Mechanics: Theory and
  Experiment}\ }\textbf {\bibinfo {volume} {2013}},\ \bibinfo {pages} {P09005}
  (\bibinfo {year} {2013})}\BibitemShut {NoStop}%
\bibitem [{\citenamefont {Lim}\ and\ \citenamefont {Sheng}(2016)}]{lim_sheng}%
  \BibitemOpen
  \bibfield  {author} {\bibinfo {author} {\bibfnamefont {S.~P.}\ \bibnamefont
  {Lim}}\ and\ \bibinfo {author} {\bibfnamefont {D.~N.}\ \bibnamefont
  {Sheng}},\ }\href {\doibase 10.1103/PhysRevB.94.045111} {\bibfield  {journal}
  {\bibinfo  {journal} {Phys. Rev. B}\ }\textbf {\bibinfo {volume} {94}},\
  \bibinfo {pages} {045111} (\bibinfo {year} {2016})}\BibitemShut {NoStop}%
\bibitem [{\citenamefont {Luitz}(2016)}]{luitz_entropy_tails}%
  \BibitemOpen
  \bibfield  {author} {\bibinfo {author} {\bibfnamefont {D.~J.}\ \bibnamefont
  {Luitz}},\ }\href {\doibase 10.1103/PhysRevB.93.134201} {\bibfield  {journal}
  {\bibinfo  {journal} {Phys. Rev. B}\ }\textbf {\bibinfo {volume} {93}},\
  \bibinfo {pages} {134201} (\bibinfo {year} {2016})}\BibitemShut {NoStop}%
\bibitem [{\citenamefont {Sahu}\ \emph {et~al.}(2019)\citenamefont {Sahu},
  \citenamefont {Xu},\ and\ \citenamefont {Swingle}}]{sahu_scrambling}%
  \BibitemOpen
  \bibfield  {author} {\bibinfo {author} {\bibfnamefont {S.}~\bibnamefont
  {Sahu}}, \bibinfo {author} {\bibfnamefont {S.}~\bibnamefont {Xu}}, \ and\
  \bibinfo {author} {\bibfnamefont {B.}~\bibnamefont {Swingle}},\ }\href
  {\doibase 10.1103/PhysRevLett.123.165902} {\bibfield  {journal} {\bibinfo
  {journal} {Phys. Rev. Lett.}\ }\textbf {\bibinfo {volume} {123}},\ \bibinfo
  {pages} {165902} (\bibinfo {year} {2019})}\BibitemShut {NoStop}%
\bibitem [{\citenamefont {Kraus}\ and\ \citenamefont
  {Cirac}(2001)}]{kraus_cirac}%
  \BibitemOpen
  \bibfield  {author} {\bibinfo {author} {\bibfnamefont {B.}~\bibnamefont
  {Kraus}}\ and\ \bibinfo {author} {\bibfnamefont {J.~I.}\ \bibnamefont
  {Cirac}},\ }\href {\doibase 10.1103/PhysRevA.63.062309} {\bibfield  {journal}
  {\bibinfo  {journal} {Phys. Rev. A}\ }\textbf {\bibinfo {volume} {63}},\
  \bibinfo {pages} {062309} (\bibinfo {year} {2001})}\BibitemShut {NoStop}%
\bibitem [{\citenamefont {Lezama}\ and\ \citenamefont
  {Luitz}(2019)}]{lezema_luitz}%
  \BibitemOpen
  \bibfield  {author} {\bibinfo {author} {\bibfnamefont {T.~L.~M.}\
  \bibnamefont {Lezama}}\ and\ \bibinfo {author} {\bibfnamefont {D.~J.}\
  \bibnamefont {Luitz}},\ }\href {\doibase 10.1103/PhysRevResearch.1.033067}
  {\bibfield  {journal} {\bibinfo  {journal} {Phys. Rev. Research}\ }\textbf
  {\bibinfo {volume} {1}},\ \bibinfo {pages} {033067} (\bibinfo {year}
  {2019})}\BibitemShut {NoStop}%
\bibitem [{\citenamefont {Majumdar}\ \emph {et~al.}(2019)\citenamefont
  {Majumdar}, \citenamefont {Pal},\ and\ \citenamefont
  {Schehr}}]{majumdar_extreme_value}%
  \BibitemOpen
  \bibfield  {author} {\bibinfo {author} {\bibfnamefont {S.~N.}\ \bibnamefont
  {Majumdar}}, \bibinfo {author} {\bibfnamefont {A.}~\bibnamefont {Pal}}, \
  and\ \bibinfo {author} {\bibfnamefont {G.}~\bibnamefont {Schehr}},\ }\href
  {http://www.sciencedirect.com/science/article/pii/S0370157319303291}
  {\bibfield  {journal} {\bibinfo  {journal} {Physics Reports}\ } (\bibinfo
  {year} {2019})}\BibitemShut {NoStop}%
\bibitem [{\citenamefont {Soeda}\ \emph {et~al.}(2014)\citenamefont {Soeda},
  \citenamefont {Akibue},\ and\ \citenamefont {Murao}}]{soeda_kc_number}%
  \BibitemOpen
  \bibfield  {author} {\bibinfo {author} {\bibfnamefont {A.}~\bibnamefont
  {Soeda}}, \bibinfo {author} {\bibfnamefont {S.}~\bibnamefont {Akibue}}, \
  and\ \bibinfo {author} {\bibfnamefont {M.}~\bibnamefont {Murao}},\ }\href
  {\doibase 10.1088/1751-8113/47/42/424036} {\bibfield  {journal} {\bibinfo
  {journal} {Journal of Physics A: Mathematical and Theoretical}\ }\textbf
  {\bibinfo {volume} {47}},\ \bibinfo {pages} {424036} (\bibinfo {year}
  {2014})}\BibitemShut {NoStop}%
\end{thebibliography}%
\end{document}